\documentclass[aps,twocolumn,showpacs,floatfix,nofootinbib,preprintnumbers,superscriptaddress,amsmath,amssymb]{revtex4-1}

\usepackage{graphicx}
\usepackage{epsfig}					
\usepackage{bm}
\usepackage{color}
\usepackage{float}
\usepackage{dcolumn}
\usepackage{subfigure}
\usepackage{multirow} 
\usepackage{amsmath}

\newcommand{\GLL}{_\mathrm{WV}}
\newcommand{\rhoWV}{\rho_\mathrm{WV}}

\begin{document}

\title{Microscopic modeling of contact formation between confined surfaces in solution} 

\author{J\o rgen H\o gberget} 
\affiliation{Department of Physics, University of Oslo, N-0316 Oslo, Norway} 

\author{Anja R\o yne} 
\affiliation{Department of Physics, University of Oslo, N-0316 Oslo, Norway}

\author{Dag K. Dysthe}
\affiliation{Department of Physics, University of Oslo, N-0316 Oslo, Norway}

\author{Espen Jettestuen} 
\affiliation{IRIS AS, P.O. Box 8046, N-4068 Stavanger, Norway}
\affiliation{Department of Physics, University of Oslo, N-0316 Oslo, Norway}				

\begin{abstract} 
We derive a Kinetic Monte Carlo model for studying how contacts form between confined surfaces in an ideal solution. 
The model incorporates repulsive and attractive surface-surface forces between
a periodic (2+1)-dimensional solid-on-solid (SOS) crystal surface and a confining flat surface.
The repulsive interaction is derived from the theory of electric double-layers, 
and the attractive interactions are Van der Waals interactions between particles on the SOS surface
and the confining surface. 
The confinement is induced by a constant external pressure
normal to the surfaces which is in mechanical equilibrium with the surface-surface forces.
The system is in thermal equilibrium, and particles can deposit to and dissolve from the SOS surface. 
The size of stable contacts formed between the surfaces in chemical equilibrium show a 
non-trivial dependency on the external pressure which is phenomenologically similar to the dependency 
of oscillatory hydration forces on the surface-surface separation. 
As contacts form we find classical phenomena such as Ostwald ripening, coalescence,
and primary and secondary nucleation stages.
We find contacts shaped as islands, bands or pits, depending solely on the contact size relative to the system size.
We also find the model to behave well out of chemical equilibrium.
The model is relevant for understanding processes where the force of crystallization and pressure solution
are key mechanisms.
\end{abstract}


\maketitle

\section{Introduction}
\label{sec:Intro}

Solids brought into contact are ubiqutous and dynamic processes and
solid contacts are central to tribology~\cite{Fischer1988,Kim2012} and the nature of granular
materials in general~\cite{Shafer1996}.
Physicists often idealize the contact dynamics 
and study inert surfaces that deform only mechanically,
since state-of-the art 
surface measurement techniques fail to work 
in the confined environments where the chemical 
reactions at the interfaces are important~\cite{Kim2012},
and the vast number of simultaneously occurring 
chemo-mechanical phenomena 
that depend on the contact topology and stresses
makes modeling difficult~\cite{Fischer1988}.
Hence the dynamics of {\em reactive} solid contacts, 
has not received the attention it deserves.
 
Reactive contact dynamics
have important applications in processes such as sintering~\cite{Orru2009, Ruths2004},
where mineral grains stick together after e.g.~compaction without liquification occurring at the grain boundaries, 
fracture healing/crack sealing~\cite{Renard2000, Fuenkajorn2011}, where material in voids and cracks is rearrange in time such that the aperture decreases
without the need of a supersaturated solution, 
the weathering of rocks and concrete~\cite{Rijniers2005, Gratier2012, Schiro2012, Flatt2014, Desarnaud2015},
which is of fundamental interest to building conservation, 
in addition to metamorphism, diagenesis and weathering in the Earth's crust~\cite{Putnis2002, Gratier2013}.

Biological applications 
span from the development of the fracture callus in the reparative stage of bone fracture healing~\cite{M2015} 
to the initial stages of cell membrane fusion~\cite{Israelachvili2011i}.

Stress-induced instabilities in reactive solid-solid boundaries are also responsible 
for the formation and evolution of stylolites~\cite{Schmittbuhl2004, Angheluta2008},
and recent experimental~\cite{ShmuelM.Rubinstein2004, Li2011} and theoretical~\cite{Tromborg2014, Thogersen2014, Srinivasan2009, Filippov2004}
studies of frictional interfaces show that the behavior of the microjunctions (i.e.~contact points) between the surfaces is crucial when determining the frictional dynamics.

Popular models of how the grain boundary
behaves during pressure solution~\cite{DenBrok1998} are growth and dissolution with the presence of a confined thin yet stable liquid film~\cite{Weyl1959},
and growth causing stabilizing island-shaped contacts between the surfaces~\cite{Raj1981}. 
Recent experiments have shown that growth rims during experiments on the force of crystallization~\cite{Royne2012},
and grains that have undergone pressure solution creep~\cite{Dysthe2002},
have a structured roughness. This is in disagreement with the liquid film model which predict smooth interfaces~\cite{Weyl1959, Royne2012}.
We therefore hypothesize that attractive surface-surface interactions are important mechanisms in confined crystallization.  

We have previously reported on a model without attractive surface-surface forces, 
which reproduce known thermodynamics for confined surfaces in solution,
as well as the pressure solution and the force of crystallization phenomena~\cite{Hogberget2016}.
However, this model also predicts a smooth interface. 
We will therefore in this work address the question whether extending the earlier model by adding a Van der Waals-like interaction between the two surfaces 
is sufficient to produce a structured roughness.

Questions we want to address are under which conditions we can expect stable contacts to form between the surfaces,
how these equilibrate and how they appear once equilibrated, and whether these contacts remain stable 
in systems where the confining surface is displaced due to the force of crystallization or the crystal surface is dissolved due to pressure solution.

The paper is structured as follows: In Sec.~\ref{sec:Model} 
we introduce all aspects of the model such as the different surfaces and the solution,
the allowed transitions and their rates, and the interactions used
and how their resulting forces are used to maintain mechanical equilibrium.
In Sec.~\ref{sec:Results} we present the results of 
how the system equilibrates, 
how the equilibrium contacts behave,
the contact fluctuations, and finally the 
out-of-equilibrium properties.
The final discussions and conclusions in Sec.~\ref{sec:Discussions} concludes the paper.

\section{Model}
\label{sec:Model}

The model consists of a periodic crystal surface placed in an ideal solution confined vertically by a flat inert surface of the same material
at a height $h_l(t)\in \mathbb{R}$.
The latter will from here on be referred to as the confining surface. 
The crystal surface is modeled using a ($2 + 1$)-dimensional periodic solid-on-solid (SOS) surface.

The SOS condition does not allow for overhangs,
hence the surface is described by an array of heights $h_i\in \mathbb{Z}$ (in units of bond lengths $l_0$), 
where $i\in[0, L\times W]$ with $L$ and $W$ being the spatial extents of the system. 
The top-most particles of the crystal surface, from here on referred to as a surface particles,
are the only ones that can take part in transitions.
The reactive surface area of the crystal is thus $A=L\times W$.
The confining surface is kept inert to reduce the complexity the model,
since SOS models are not applicable to systems where two opposing surfaces fluctuate in and out of contact with each other (formed contacts could never break).

The liquid surrounding the crystal has a uniform concentration of solute particles,
which limits the model to reaction limited systems. Adding a more realistic description 
of the liquid is possible, but we will here keep the description as simple as possible. 
The concentration level may vary in time.

Allowed transitions in the system are dissolution of crystal surface particles into solution 
and deposition of solute particles to the crystal surface. 

\begin{figure*}[t]
 \begin{center}
\includegraphics[width=0.9\textwidth]{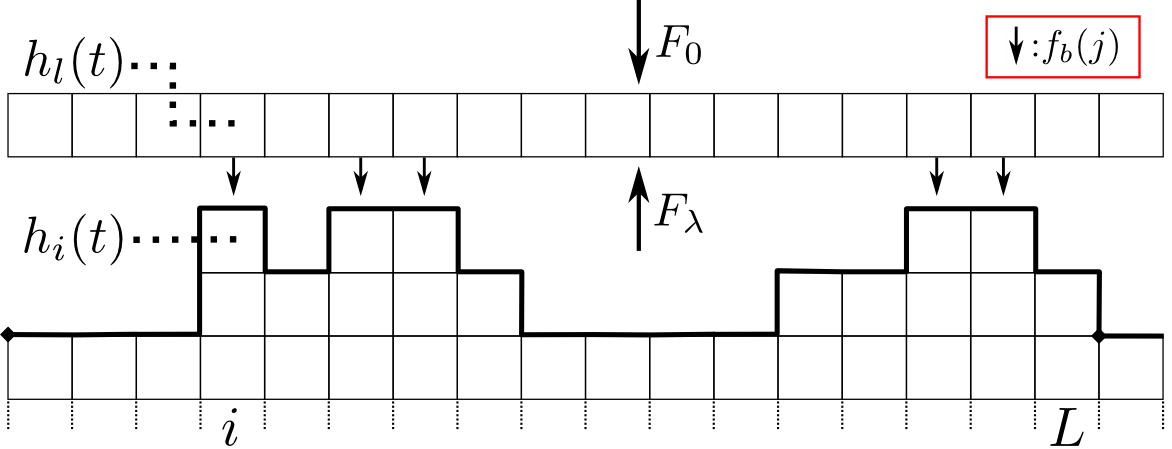}
 \end{center}
 \caption{An illustration of a 1-dimensional slice of the surfaces contained in the model. The top surface, referred to as the confining surface,
 is inert, perfectly plane and placed at a position $h_l(t)\in\mathbb{R}$,
 and the bottom surface (thick line) is a periodic solid-on-solid surface (no overhangs) made up of $L$ heights $h_i(t)\in \mathbb{Z}$.
 The arrows indicate forces acting on the confining surface, which are a constant external force $F_0$, a surface-surface repulsive force $F_\lambda$, and 
 attractive particle-particle Van der Waals-like forces $f_b$. The confining surface height $h_l(t)$ is set such that these forces are in equilibrium, or
 the surfaces are resting on one another. Between the surfaces there is an uniform ideal solution at a given concentration level.}
 \label{fig:system}
\end{figure*}

Particles interact with other particles through nearest neighbor interactions with bond energy $E_b$.
The confining surface is subject to an external force with magnitude $F_0$ in the direction normal to the confining surface. 
A repulsive force $F_\lambda$ is generated between the surfaces which increases as the separation decreases.
This far the model is identical to the one used in our earlier work on the effect of normal stress on confined crystals \cite{Hogberget2016}.
Here we will include additional short-range attractive forces $f_b(i)$ between the surface particles and the confining surface.
We enforce the mechanical equilibrium of the confining surface, that is, 
the repulsive force always balances the external- and attractive forces.
This involves repositioning the confining surface height $h_l(t)$.
The acting forces are illustrated together with the surfaces in Fig.~\ref{fig:system}.

For an initial volume $V(0)$ and concentration $c(0)$, the effective number of solute particles is $N_\mathrm{s}(0) = c(0)V(0)$.
Since the system is periodic particles have no means of escaping or entering the system,
which means that the total number of particles is conserved. 
We can therefore keep track of $N_\mathrm{s}(t)$ by counting the number of dissolved and deposited particles and add it to the initial value. 
Hence the concentration at a time $t$ is $c(t) = N_\mathrm{s}(t)/V(t)$.

\subsection{State Transitions and Rates}
\label{sec:state_trans_and_rates}

A deposition can occur at any site given that the confining surface does not block it.
A dissolution can occur at any site given that there is an available neighboring site.
These restrictions on the deposition and dissolution reactions ensures that the surfaces do not penetrate into one another.

If the neighboring dissolution site is in the solution, the surface particle dissolves and is removed from the surface. 
If on the other hand the neighboring site is at the crystal surface, the particle slides one lattice length horizontally.
This can be interpreted as an immediate dissolution-deposition chain, and is important to include
in order to avoid surface particles in regions where the surfaces are in contact becoming static.
Horizontal sliding is the only surface-surface transition we allow, since including transitions up or down kink sites in a SOS model
is known to cause an anisotropy between vertical and horizontal diffusion~\cite{Kotrla1996, Petrov2014}. 

We use that the rate of a particle $i$ dissolving is~\cite{Hogberget2016}

\begin{equation}
\label{eq:escaperate_orig}
 R_-(i) = \nu \exp\boldsymbol{(}-\Delta G(i)/kT\boldsymbol{)},
\end{equation}

\noindent
where $\nu$ is a frequency factor, $\Delta G(i)$ is the free energy gain by removing particle $i$ from the system, 
$k$ is the Boltzmann constant and $T$ is the temperature. 

The deposition rate is proportional to the current concentration $c$ as follows:

\begin{equation}
 R_+(i) = \nu c,
\end{equation}

\noindent
where we for simplicity have used the same frequency factor such that it can be used to set the time scale for the simulations.
The system is in chemical equilibrium when $c$ has an equilibrium value $c_\mathrm{eq}$
at which no net growth occurs. Using a different frequency $\nu_+$ would cause the system to equilibrate 
at a different concentration $\tilde c_\mathrm{eq} = c_\mathrm{eq} \nu/\nu_+$.

\subsection{Free Energies}

We model $\Delta G(i)$ from Eq.~(\ref{eq:escaperate_orig}) as three terms representing three interactions as follows:

\begin{equation}
\label{eq:delta_g_for_rate}
 \Delta G(i) = E_b n_i  + \Delta G\GLL(i) + \Delta G_\lambda(i),
\end{equation}

\noindent
where the first term is the particle-particle interaction represented by the breaking of $n_i$ nearest neighbor bonds with energy $E_b$,
the second term represents the particle-surface attraction, and the last term represents the surface-surface repulsion.

We set the following criteria for the free energies, which determines the shape and interplay of the different interactions:

\begin{enumerate}
 \item When in perfect contact (no liquid present), the two surfaces produce an infinite bulk structure consisting only of nearest neighbor bonds.
 \item The surfaces are made up of the same material.
 \item The attractive particle-particle interactions has a Van der Waals-like decay as $1/d^6$~\cite{Israelachvili2011i}, 
 where $d \ge 1$ is the center-center separation (in lattice units).
\end{enumerate}

The free energy due to the external force $F_0$ is $F_0 h_l$, where $h_l$ is the height of the confining surface.
This is analogous to the gravitational potential. This free energy does not directly
depend on the crystal surface structure, hence there is no change in free energy associated with $F_0$
present in the rate calculations.

\subsubsection{Surface-surface repulsion}

We use the same expression for the repulsive interaction as in Ref.~\cite{Hogberget2016}
with the exception that it goes to $0$ for no separation:

\begin{equation}
\label{eq:dg_lambda}
 \Delta G_\lambda(i)/E_b = \left\{\begin{array}{clc}
        -\sigma_0\exp\boldsymbol{(}-(d_i-1)/\lambda_D\boldsymbol{)} &,&  d_i > 1\\
         0 &,& \text {else}
        \end{array}\right.,
\end{equation}

\noindent
where $d_i\ge 1$ is the center-center distance between the outer particles of the confining surface and surface particle $i$ in lattice units. 
The free parameter $\sigma_0$ represents the ratio between the strength of the surface-surface repulsion and the particle-particle attraction $E_b$.
The decay strength $\lambda_D$ is analogous to the Debye length.
We will use $\lambda_D = 5$ throughout this work and vary the value of $\sigma_0$ to study the effect of an increasing/decreasing repulsion.

The reason why we set the interaction to $0$ for $d_i=1$ is that we assume that the repulsion originates from the existence of an electrolyte,
and if there is no room for liquid, the only interaction is single bonds with energy $E_b$. 
Moreover, keeping the repulsion active for $d_i=1$ would violate the first criteria listed earlier concerning the convergence to a bulk material. 
The decay to 0 repulsion is probably smooth, and the abrupt cut used here is a simplification.
This simplification does not cause instabilities since if $d_i=1$ at any point,
we set the total force to 0 (balanced by a normal force).
Hence forces at $d_i=1$ does not need to be evaluated explicitly.

The total free energy due to this interaction is 

\begin{equation}
\label{eq:G_lambda}
 G_\lambda/E_b = -\frac{1}{\zeta}\sum_i \Delta G_\lambda(i)/E_b,
\end{equation}

\noindent
where $\zeta \equiv 1 - \exp(-1/\lambda_D).$

\subsubsection{Particle-particle attraction}

\begin{figure}[t]
 \begin{center}
  \includegraphics[width=0.4\paperwidth]{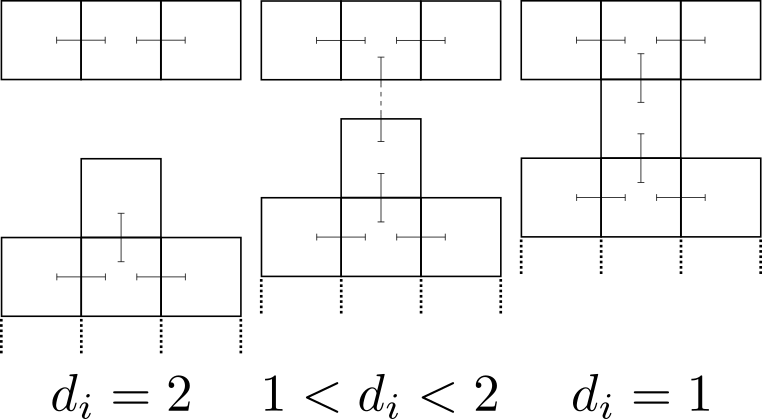}
  \caption{Illustrations of the three cases in Eq.~(\ref{eq:Gl_conditions}) concerning the free energy between a particle on the
  crystal surface and a particle in the confining surface separated by a distance $d_i$ in lattice units. The main idea is that this interaction behaves like the nearest neighbor
  interaction for integer separations. This is illustrated by $d_i=2$ (left) and $d_i=1$ (right) having no bond and a nearest neighbor bond, respectively.
  The interaction at intermediate separations (middle), indicated by a stippled bond, is not described by the nearest neighbor model.
  We model this interaction as a $1/d_i^6$ Van der Waals-like decay, with added corrections to ensure a continuous transition to 0 for $d_i=2$ for both the function and its derivative. 
  The interaction is given by Eq.~(\ref{eq:dg_pretty}).}
  \label{fig:attraction_illustrated}
 \end{center}
\end{figure}

The surface particles are confined to a lattice and thus always have a bond length separation and consequently share an energy $E_b$ if they are
nearest neighbors, and 0 else. The position of the confining surface, however, is not restricted to integer bond lengths, 
hence intermediate separations $d_i\in(1,2)$, that is, separations between the nearest and next nearest neighbor, 
have some value in $G\GLL(i)\in(0, E_b)$ which we assume decays as $1/d_i^6$ (Van der Waals attraction).

Summarized we may write

\begin{equation}
\label{eq:Gl_conditions}
 \tilde G\GLL(i) = \left\{\begin{array}{clc}
        -E_b &,&  d_i = 1\\
        G\GLL(i) &,& d_i\in(1,2) \\
        0 &,&  \text {else}
        \end{array}\right.,
\end{equation}

\noindent
This attraction is illustrated for all three cases in Fig.~(\ref{fig:attraction_illustrated}).

\begin{figure}[t]
 \begin{center}
   \includegraphics[width=0.4\paperwidth]{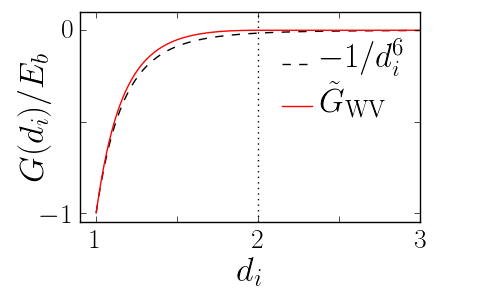}
  \caption{Comparison between the attractive interaction used in this work $\tilde G\GLL(d_i)$,
  and the Van der Waals-like decaying function $-1/d_i^6$ it is designed to resemble, where $d_i$ is the surface-surface separation at site $i$
  in lattice units. 
  It is evident that there is indeed a strong resemblance.
  The reason for the difference is that the attractive interaction, as well as its derivative, is required to be 0 at a separation of $d_i=2$ (stippled vertical line). 
  This requirement is a consequence of our assumption that this attractive interaction should act identical to the nearest neighbor interaction used on the crystal surface 
  for integer separations. This is also the reason why for separations $d_i=1$ we have a free energy equal to the bond energy $E_b$.}
  \label{fig:wv_compare_6}
 \end{center}
\end{figure}

We cannot choose $G\GLL(i) = -E_b/d_i^6$ directly, since this would render the free energy discontinuous at $d_i=2$.
We must therefore correct the potential with a shift (as is often done with the Lennard-Jones interaction \cite{Frenkel200223}).
A constant shift, however, would still leave the derivative of the free energy discontinuous at $d_i=2$, hence
we introduce a shift in the derivative as well (a linear term in the original expression).
The expression now reads

\begin{equation}
\label{eq:dg_pretty}
   G\GLL(i)/E_b = -\frac{1}{60}(3d_i + 64/d_i^6  - 7),
\end{equation}

\noindent
which when inserting $d_i=1$ and $d_i=2$ produce the correct values $E_b$ and 0 such that Eq.~(\ref{eq:Gl_conditions})
is continuous. In Fig.~\ref{fig:wv_compare_6} we have compared this expression to the $~1/d_i^6$ function it was designed to resemble,
and it is evident that this is in fact the case.

The derivative of Eq.~(\ref{eq:dg_pretty}) with respect to $d_i$ is
\begin{equation}
\label{eq:gl_derivative}
 \frac{\partial}{\partial d_i}  G\GLL(i)/E_b = \frac{1}{20}\left(128/d_i^7 - 1\right),
\end{equation}

\noindent
which equals 0 when $d_i = 2$, which means that the derivative of Eq.~(\ref{eq:Gl_conditions})
is continuous as required.

When we remove a surface particle $i$ with a separation $d_i\in[1, 2)$, that is, 
the particle has a non-zero attractive interaction with the confining surface, we are guaranteed that the new surface particle $i$ 
has $d_i\in[2, 3)$ and consequently no attraction.
The change in free energy used in Eq.~(\ref{eq:delta_g_for_rate}) is therefore 

\begin{equation}
\label{eq:gl_eq_dgl}
 \Delta G\GLL(i) = 0-G\GLL(i) = \frac{1}{60}(3d_i + 64/d_i^6  - 7).
\end{equation}

\subsection{Mechanical equilibrium}
\label{sec:mech_eq}

\begin{figure}
 \includegraphics[width=0.4\paperwidth]{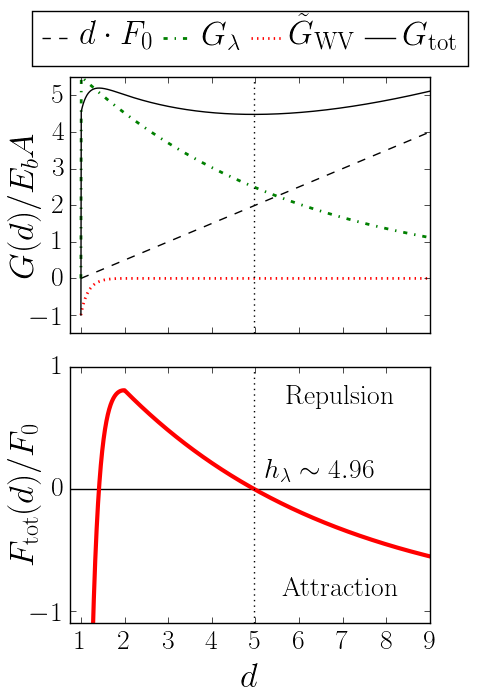}
 \caption{A visualization of the separation ($d$) dependency of the different free energy sources available in the model which affects the confining surface (top panel)
 and the resulting total force on the confining surface (bottom panel) for a perfectly flat crystal surface using a repulsion strength $\sigma_0 = 1$ 
 relative to the bond energy $E_b$, and an external force with magnitude $F_0/E_bA=0.5$. The potential from the constant external force becomes $F_0 d$ similar to that of 
 a gravitational potential, the attractive interaction $\tilde G\GLL$ is shown in more detail in Fig.~\ref{fig:wv_compare_6}, 
 and the free energy due to repulsion is given by Eq.~(\ref{eq:G_lambda}). We observe multiple equilibria:
 one where the surfaces are resting on each other ($d=1$), one peak in the free energy at a short separation where 
 a strong attraction is in equilibrium with a strong repulsion, 
 and one valley where a weaker (or constant) attraction is in equilibrium with a weaker repulsion.
 We do not include transitions to the unstable peak equilibrium, since we allow 
 only transitions between stable states. The challenge is thus obtaining the second stable equilibrium here located at $h_\lambda\sim 4.96$
 indicated by the vertical stippled line. The force peak will appear smoother for rough surfaces.}
 \label{fig:free_energies_and_force}
\end{figure}

\begin{figure}
 \begin{center}
  \includegraphics[width=0.4\paperwidth]{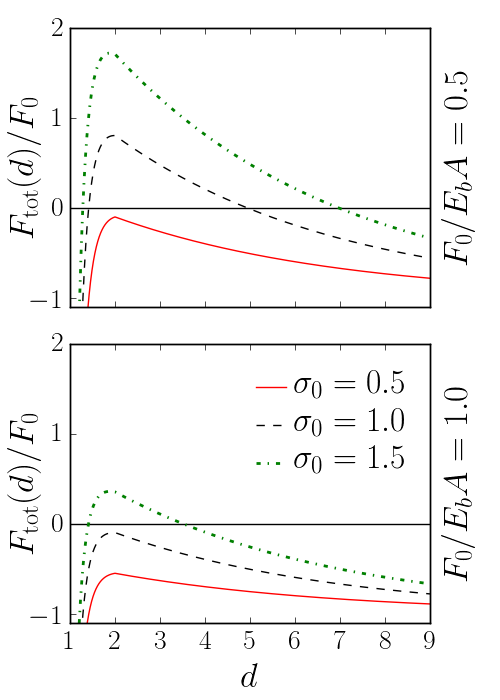}
 \caption{The total force $F_\mathrm{tot}(d)$ on the confining surface at a separation $d$ from a flat crystal surface, 
 where $F_0/E_bA$ is the unit less pressure generated by the external force of magnitude $F_0$,
 $\sigma_0$ is the magnitude of the repulsive interaction relative to the binding energy $E_b$ (attractive interaction).
  We observe that as we make the repulsion weaker ($\sigma_0$ smaller), the equilibria disappear since the attractive
  interaction overcomes the repulsive interaction before it comes out of range at $d=2$. In these cases the only
  valid equilibrium is when the two surfaces are resting on each other, 
  i.e.~the surfaces snap into contact due to an overwhelming attraction. A flat surface under these conditions
  will always be in contact, however, an arbitrary rough surface can produce many different force profiles under the same conditions.
  Nevertheless, these examples give an important insight into how we can expect the force profile to look, and what it means for the different
  equilibria.}
 \label{fig:forces_all_params}
  \end{center}
\end{figure}

The forces on the confining surface is balanced at all times.
This gives the equation

\begin{equation}
\label{eq:mechanical_equilibrium_orig}
 F_\lambda + \sum_j f_b(j) - F_0 = 0,
\end{equation}

\noindent
where $F_\lambda \ge 0$ is the generated repulsive force, $f_b(j) \le 0$ is the attractive interaction at surface site $j$, and $F_0 > 0$ is the magnitude of the external force. 
These are illustrated in Fig.~\ref{fig:system}. 

If the two surfaces are resting on one another, i.e., $d_i=1$ for any $i$, the net force is always 0 due to a normal force associated with the contact.
This is a necessary criterion since we do not model the repulsive forces for $d_i<1$,
which could have been included by adding a $1/d_i^{12}$ term, i.e.~use the Lennard-Jones 12-6 potential~\cite{Frenkel200223} instead of a pure Van der Waals attraction.

The height at which the confining surface is resting on the crystal surface, $h_c$, is related to the maximum surface height as $h_c = \max(\textbf{h}) + 1$, 
where $\textbf{h}$ denotes the array of all surface heights $h_i$.

The force is calculated as $F=-\partial G/\partial h_l$.
The calculation of the repulsive force is straight forward:

\begin{equation}
 F_\lambda = \frac{1}{\lambda_D} G_\lambda,
\end{equation}

\noindent
where we have set the force in contact equal to zero, which is necessary since the potential is discontinuous due to our abrupt cutting of the interaction potential.
Writing the separation as $d_i = h_l - h_i$, it is clear that $\partial d_i / \partial h_l = 1$, such that differentiating with respect to $d_i$ or $h_l$ yields the same result. 
Using this, the attractive forces become

\begin{equation}
 f_b(i) = \left\{\begin{array}{clc}
        -\partial  G\GLL(i)/\partial d_i &,& d_i\in(1,2) \\
        0 &,&  \text {else}
        \end{array}\right.,
\end{equation}

\noindent
where the derivative is given in Eq.~(\ref{eq:gl_derivative}).
When $d_i=1$ for any $i$, i.e. the surface is resting on one another, we assume that the total force is zero.
Hence we do not have to be concerned about instabilities at $d_i=1$.

Figure~\ref{fig:free_energies_and_force} shows the total free energy together with its individual contributions (top panel) 
and the resulting total force (bottom panel) for a flat surface.
From the bottom panel we see two candidates for out-of-contact equilibria ($F_\mathrm{tot}(d) = 0$),
however, looking at the top panel we see that the one closest to the crystal surface is unstable.
In KMC we allow only transitions to stable states, hence we only consider the far equilibrium point $h_\lambda$ and the contact point at $d=1$.

Maintaining mechanical equilibrium thus boils down to calculating $h_\lambda$ and deciding whether to choose it or the contact height $h_c$.
We split this into two cases: when $h_l=h_c$, i.e.~when the surfaces are resting on one another, and when they are not.

When the surfaces are not resting on one another, the confining surface is moved to the closest equilibrium height ($h_c$ or $h_\lambda$) in the direction of the current total force.
Given that the position of the confining surface is $h_l$, and the maximum of the force (the peak in Fig.~\ref{fig:free_energies_and_force}) is located at $h_m$, 
then this rule translates into the following conditions: 

\begin{enumerate}
 \item $F_\mathrm{tot}(h_l) > 0:$ the surface is repelled to $h_\lambda$.
 \item $F_\mathrm{tot}(h_l) < 0:$
 \begin{enumerate}
     \item $F_\mathrm{tot}(h_m) < 0:$ $h_c$ is the only solution. 
     \item $h_l > h_\lambda:$ the surface is attracted to $h_\lambda$.
     \item $h_l < h_\lambda:$ the surface is attracted to $h_c$. 
 \end{enumerate}
\end{enumerate}

Condition 2(a) represents the case where the repulsive interaction is too weak to withstand the applied force. 
In Fig.~\ref{fig:forces_all_params} the total force for commonly used values in this paper is shown,
and we see that condition 2(a) occur for high external forces $F_0$ and/or a weak repulsive interaction strength $\sigma_0$.

If $h_\lambda>h_c+2$, we are outside the cutoff in the attractive interactions, and $h_\lambda$ has an analytical solution \cite{Hogberget2016}:

\begin{equation}
\label{eq:h_lambda_analytical}
\tilde h_\lambda = 1 + \lambda_D\log\left(\frac{\sigma_0\Theta}{\lambda_D\xi F_0/E_bA}\right),
\end{equation}

\noindent
where $\Theta = \langle\exp(h_i/\lambda_D)\rangle_i$.

In practice, we calculate this value, and if it is larger than $h_c+1$,
we know it is a valid solution. 

If the surfaces are resting on one another, i.e.~$h_l=h_c$, the net force is 0 by assumption, and is no longer responsible for the dynamics of the confining surface. 
If the contacts are unstable, dissolution is the primary mechanism for separating the surfaces, and if the contacts are stable and the system is in 
chemical equilibrium, the attraction is so strong that condition 2(a) applies almost exclusively.
However, if the solution is supersaturated, the repulsive energy in the system will steadily increase as particles deposit,
and since the repulsion has a longer range than the attractive interaction, the two equilibra may coexist even when $h_l=h_c$.
In this case we need a criterion that determines when contact bonds break and the confining surface moves from $h_c$ to $h_\lambda$.

This should be expressed in terms of a rate which depends on the energy barrier between the two states,
however, we are unable to do this since we here use rates of the form of Eq.~(\ref{eq:escaperate_orig}) which assumes single particle
transitions. In other words, if we want an implicit condition for separating the surfaces, 
we should use the Eyring rate equation~\cite{Eyring1935} directly.

Since this condition does not impact the equilibrium simulations, 
and the most important part of the separation mechanism is that condition 2(a) stops applying due to a buildup of repulsive energy, 
the actual condition is not as important as it might seem.
We therefore choose to model it based on simple thermodynamical considerations combined with an attempt frequency.

The probability that the surfaces separate in a given attempt we model as a Boltzmann weight using the change in free energy per area between the contact 
and the separated states

\begin{equation}
\label{eq:bond_break_P}
 P_b = \frac{1}{\mathcal{Z}}\exp\boldsymbol{(}-\langle \Delta G_\mathrm{tot}(h_c\to h_\lambda)\rangle/kT\boldsymbol{)} \equiv W_b/\mathcal{Z}, 
\end{equation}

\noindent
where $\mathcal{Z} = 1 + W_b$ since the weight associated with staying $W_\mathrm{stay}=1$.
When we calculate $\Delta G_\mathrm{tot}$, we do not include the repulsive 
free energy at $h_\lambda$ for the points in contact at $h_c$.
This represents a sort of retardation time needed to form the electric double layers between the newly formed surface areas.

In this paper we fix the attempt frequency to once every $A$ cycles,
which we implemented by scaling $P_b$ by $1/A$ using a frequency of 1.
This choice ensures that the probability that the surfaces separate in a given time interval 
will not have an unphysical dependency on the system size, 
since the fact that we move only one particle per cycle in KMC makes the time step $\Delta \tau_\mathrm{KMC}\propto 1/A$~\cite{Voter2007}.

\section{Results}
\label{sec:Results}

We will focus on understanding the equilibrium properties of systems in which 
stable contacts form between the surfaces. We want to understand how these contact form from an initial separated state,
which shapes they possess and why, and how they fluctuate in time.

In order to achieve this we first need a proper definition of a contact. 
Macroscopically it suffices to define a contact as a point where the surfaces rests on one another. 
Microscopically, however, we need a less binary definition, 
since the attractive interactions promoting surface-surface contacts do not necessarily require the surfaces to rest on one another to do so.
We will therefore define a contact as a point which is within the range of the attractive interaction, that is, if $h_l - h_i < 2$, then 
site $i$ is in contact. The contact density is then

\begin{equation}
\label{eq:rho_defined}
\rhoWV \equiv \frac{1}{A} \sum_i \kappa(i),
\end{equation}

\noindent
where $A$ is the area of the (flat) confining surface and $\kappa(i)$ is 1 if $h_l - h_i < 2$ and 0 else. 
We have $\rhoWV\in[0,1]$, where $\rhoWV=0$ represents completely separated surfaces, and $\rhoWV=1$ represents 
the case where all surface sites are in range of the attraction. For any realistic choices
of repulsion strength $\sigma_0$, the latter scenario leads to the surfaces joining together perfectly.

We initialize the system in an a priori known equilibrium state of the system without an attractive interaction \cite{Hogberget2016},
which is obtained by setting an initial concentration $\ln c(0) = (F_0/E_bA - 3)E_b/kT$
and the confining surface to the height given by Eq.~(\ref{eq:h_lambda_analytical}).
Note that for high pressures this initial state may have initial contacts.
We then use the method described in Section~\ref{sec:state_trans_and_rates} to keep 
a constant effective number of particles, such that the closed system will equilibrate automatically.
The initial crystal surface is random with an average height equal to 0.

All simulations are done using a $30\times 30$ surface lattice. We have done sample simulations using a $50\times 50$ surface lattice 
and observed no noticeable change in the results. The mechanical equilibrium calculations
are also very CPU-intensive, meaning that if we want to do a thorough analysis of a vast parameter space,
keeping the size as small as possible is very favorable. Periodic boundaries are also very forgiving on the system size,
so if we were to open up a boundary, caution should be taken to ensure that the boundary effects are still negligible.

\newcommand{\widtheq}{0.47}
\begin{figure*}
 \begin{center}
  \includegraphics[width=\widtheq\textwidth]{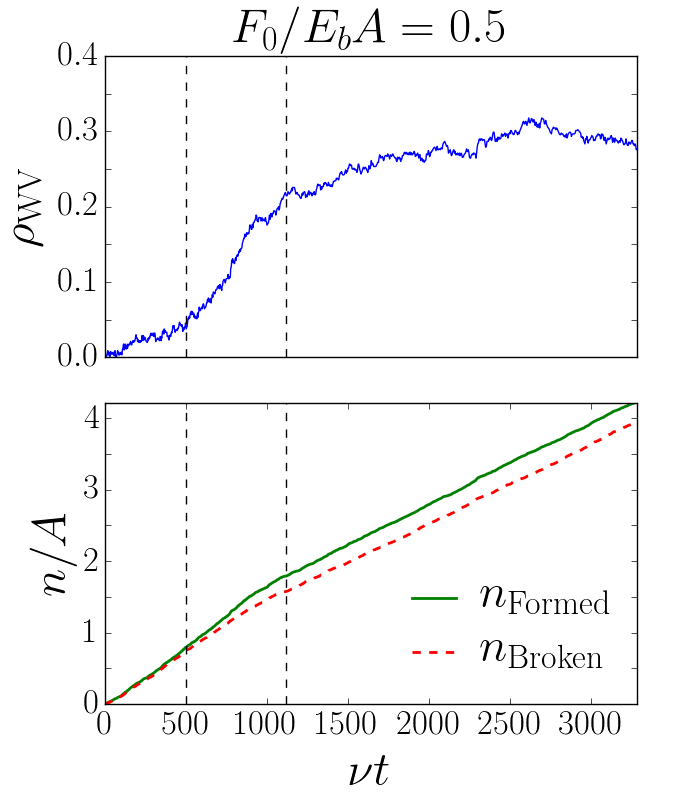} 
  \includegraphics[width=\widtheq\textwidth]{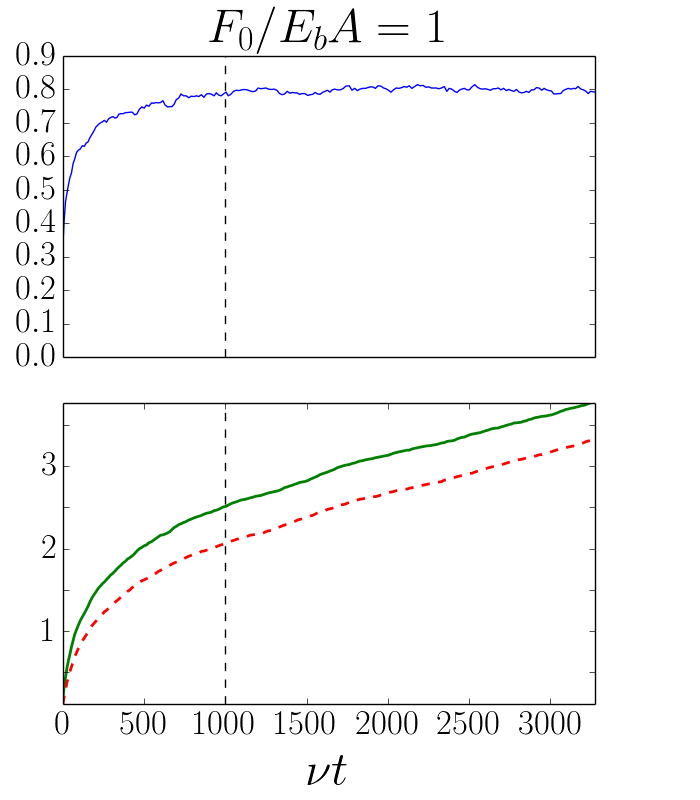} \\
  \includegraphics[width=\widtheq\textwidth]{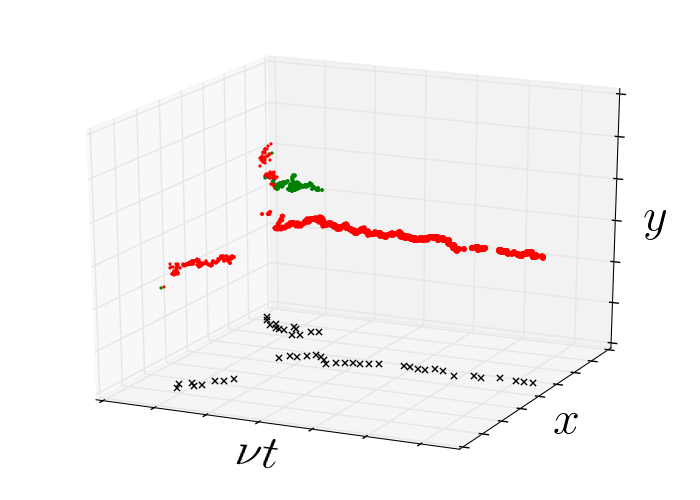}  
  \includegraphics[width=\widtheq\textwidth]{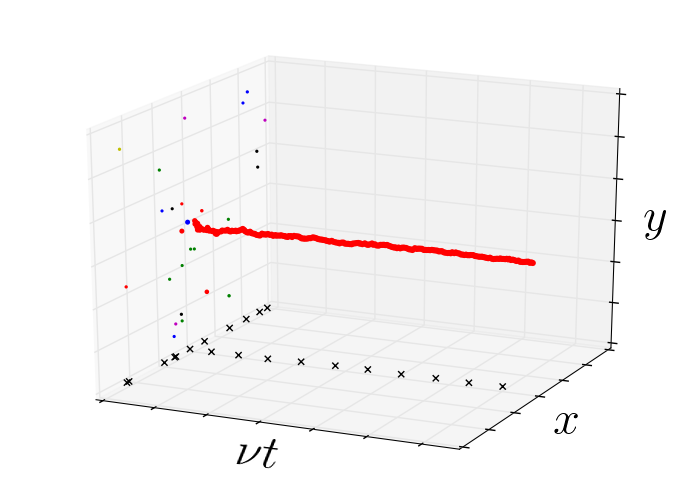} \\
 \caption{Equilibration from an initial state at time $t\nu = 0$ to a state where
 the contact density $\rhoWV$ has stabilized for two different levels of applied pressures $F_0/E_bA$
 using $\sigma_0 = 1$ and $E_b/kT=1$. The left column shows results at $F_0/E_bA = 0.5$, which is 
 considered a low pressure, and the right column shows results at $F_0/E_bA = 1$, which is considered a high pressure.
 The top row shows how the contact density grows with time. 
 For low pressures, we observe a primary stage where the 
 contacts grow slowly, before reaching a secondary stage where new contacts are formed rapidly. This stage eventually 
 ends when we have reached chemical equilibrium. The stages are separated by vertical dashed lines. 
 The second row shows the accumulated number of formed and broken contacts in the system,
 which clearly reveal the differences between the three stages just described.
 The final row shows how the contact cluster centeroids move in time. The crosses are vertical projections separated by 
 a constant number of simulation cycles (the time step is not constant). 
 For $F_0/E_bA=0.5$ we see two prominent clusters coalescing into a single cluster as the secondary stage ends.
 For $F_0/E_bA=1$, the first stage immediately ends, and smaller clusters dissolve in favor of growing one large dominant cluster.}
 \label{fig:equilibation}
 \end{center}
\end{figure*}

\subsection{Equilibration}
\label{sec:equilibration}

In this section we investigate how the contacts in the system form
as the system evolves from the initial state to its equilibrium state.

In Fig.~\ref{fig:equilibation}, we present a high and low pressure simulation of the same system.
For the low pressure case we observe three stages. In the primary stage the number of contacts increase very slowly,
since most of the contacts which form are unstable due to their small size and dissolve quickly. In the secondary stage
growth is rapid since here one or more contacts are above the critical size needed to stay stable. Finally we enter a stage
where all the contacts have coalesced into a single contact 
where the average number of formed contacts equals the average number of broken contacts.
This process has a clear similarity to kinetic limited growth theory which passes through the same stages towards the stationary state~\cite{Saito1996}.

For even lower pressures the initial surface-surface separation is too large for any stable contact clusters to appear,
and the equilibrium state becomes equal to that with no attractive interactions (total separation).
For higher pressures initial contacts are much easier to produce since the surfaces on average are closer to one another,
and often start out with parts being in contact already. The primary stage is thus skipped. 

We also observe that smaller clusters
dissolve in favor of the larger ones, which is analogous to Ostwald ripening~\cite{pimpinelli}.
This process occurs since the concentration needed to stabilize a contact cluster decreases as the size of the cluster increases
because a larger cluster is harder to dissolve.
Hence the largest contact cluster has the fastest net growth,
which makes the solution undersaturated for smaller contact clusters.
Equally sized clusters will, given enough time, coalesce into a single dominant contact.

\subsection{Equilibrium Contacts}
\label{sec:equilibrium_contacts}

\newcommand{\ifigi}{31}
\newcommand{\ifzi}{0.65}
\newcommand{\iri}{0.30}

\newcommand{\ifigii}{33}
\newcommand{\ifzii}{0.61}
\newcommand{\irii}{0.31}

\newcommand{\ifigiii}{36}
\newcommand{\ifziii}{0.77}
\newcommand{\iriii}{0.44}

\newcommand{\ifigiv}{39}
\newcommand{\ifziv}{0.74}
\newcommand{\iriv}{0.50}

\newcommand{\ifigv}{60}
\newcommand{\ifzv}{0.85}
\newcommand{\irv}{0.72}

\newcommand{\iifigi}{26}
\newcommand{\iifzi}{0.50}
\newcommand{\iiri}{0.27}

\newcommand{\iifigii}{27}
\newcommand{\iifzii}{0.59}
\newcommand{\iirii}{0.31}

\newcommand{\iifigiii}{35}
\newcommand{\iifziii}{0.65}
\newcommand{\iiriii}{0.47}

\newcommand{\iifigiv}{38}
\newcommand{\iifziv}{0.77}
\newcommand{\iiriv}{0.63}

\newcommand{\iifigv}{60}
\newcommand{\iifzv}{0.99}
\newcommand{\iirv}{0.81}

\newcommand{\iiifigi}{32prev}
\newcommand{\iiifzi}{0.60}
\newcommand{\iiiri}{0.42}

\newcommand{\iiifigii}{34} 
\newcommand{\iiifzii}{0.60} 
\newcommand{\iiirii}{0.48}

\newcommand{\iiifigiii}{38}
\newcommand{\iiifziii}{0.72}
\newcommand{\iiiriii}{0.65}

\newcommand{\iiifigiv}{40}
\newcommand{\iiifziv}{0.80}
\newcommand{\iiiriv}{0.65}

\newcommand{\iiifigv}{56}
\newcommand{\iiifzv}{0.86}
\newcommand{\iiirv}{0.82}

\newcommand{\eqswtwo}{0.22}
\newcommand{\colsepeqs}{1.5cm}
\begin{figure*}
 \begin{center}
 \begin{tabular}{rcrcr}
 $E_b/kT=0.5$ \hspace{1cm} & \hspace{\colsepeqs} & $E_b/kT=1$ \hspace{1cm} & \hspace{\colsepeqs} & $E_b/kT=2$ \hspace{1cm} \\
 \hline \\
  \includegraphics[width=\eqswtwo\textwidth]{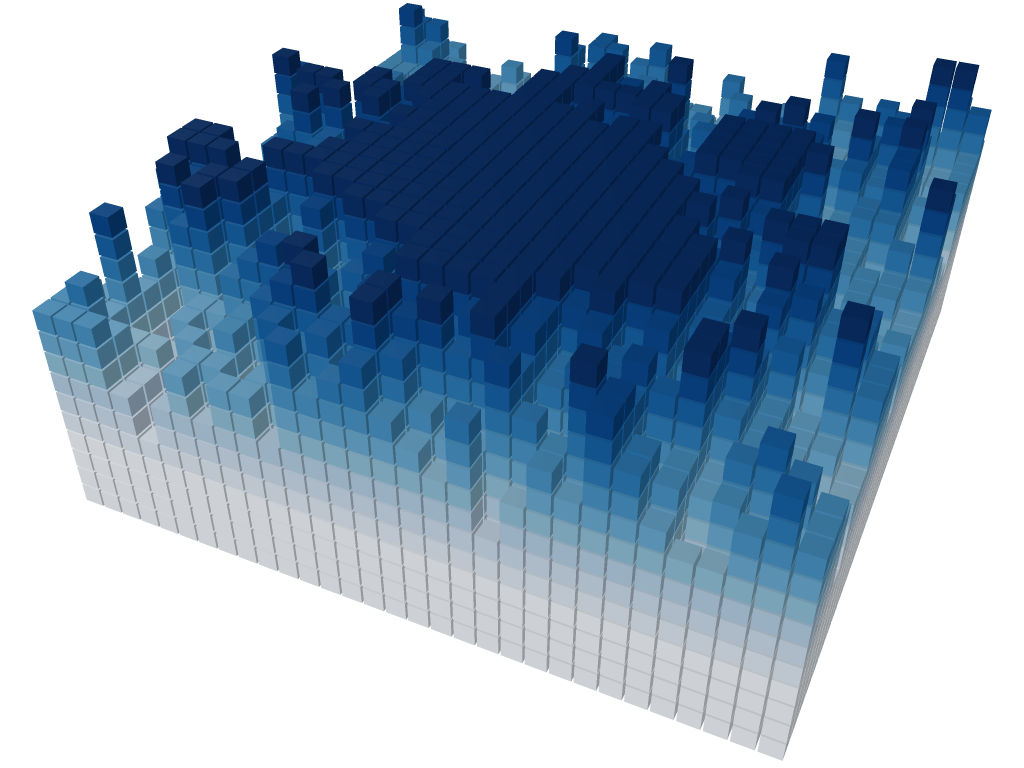} & \hspace{\colsepeqs} & 
  \includegraphics[width=\eqswtwo\textwidth]{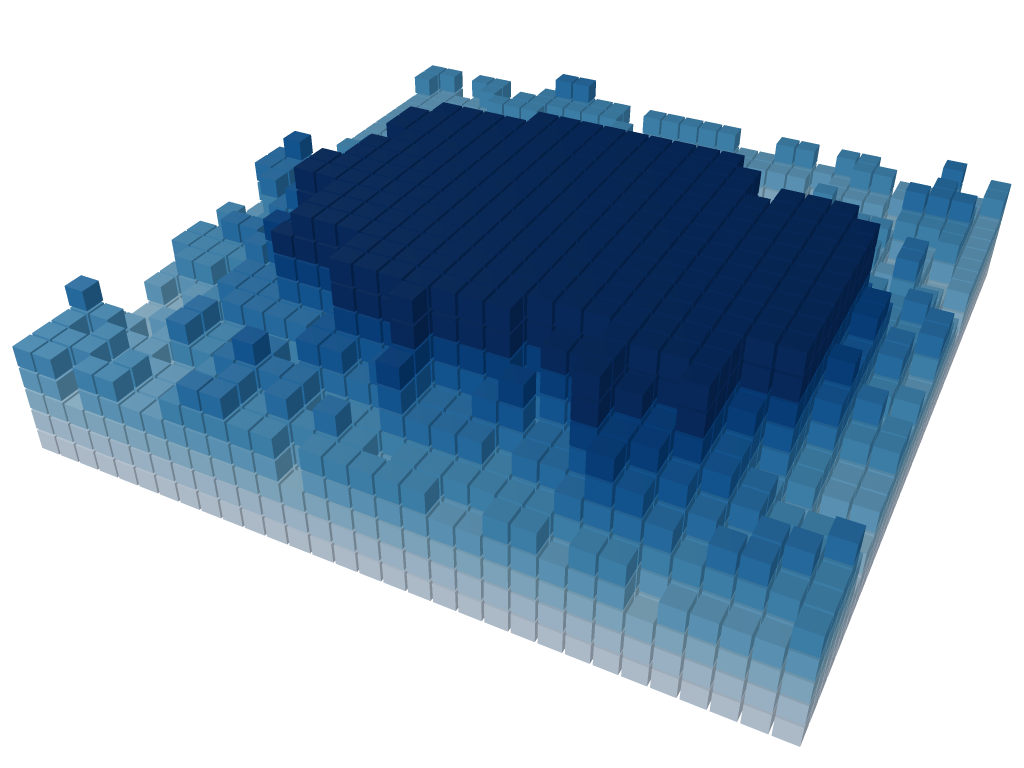} & \hspace{\colsepeqs} &
  \includegraphics[width=\eqswtwo\textwidth]{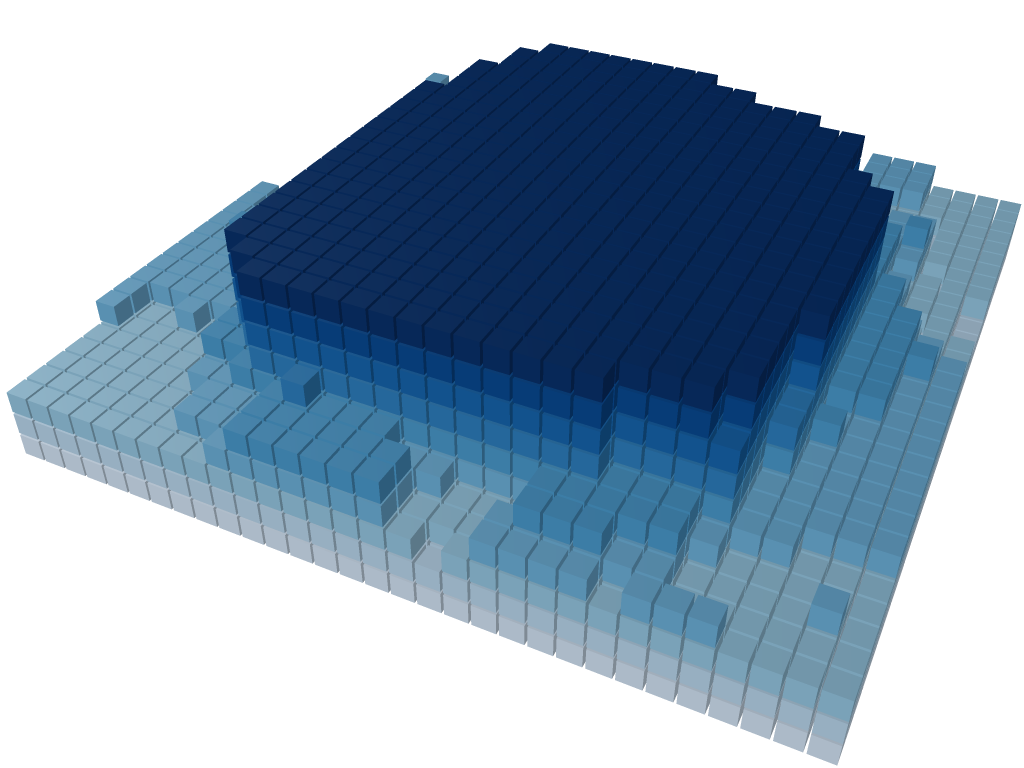} \\
  $F_0/E_bA=\ifzi$ \hspace{1cm} & \hspace{\colsepeqs} & $F_0/E_bA=\iifzi$ \hspace{1cm} & \hspace{\colsepeqs} & $F_0/E_bA=\iiifzi$ \hspace{1cm} \\
  $\rhoWV=\iri$    \hspace{1cm} & \hspace{\colsepeqs} & $\rhoWV=\iiri$    \hspace{1cm} & \hspace{\colsepeqs} & $\rhoWV=\iiiri$    \hspace{1cm} \\
  & & & & \\
  \includegraphics[width=\eqswtwo\textwidth]{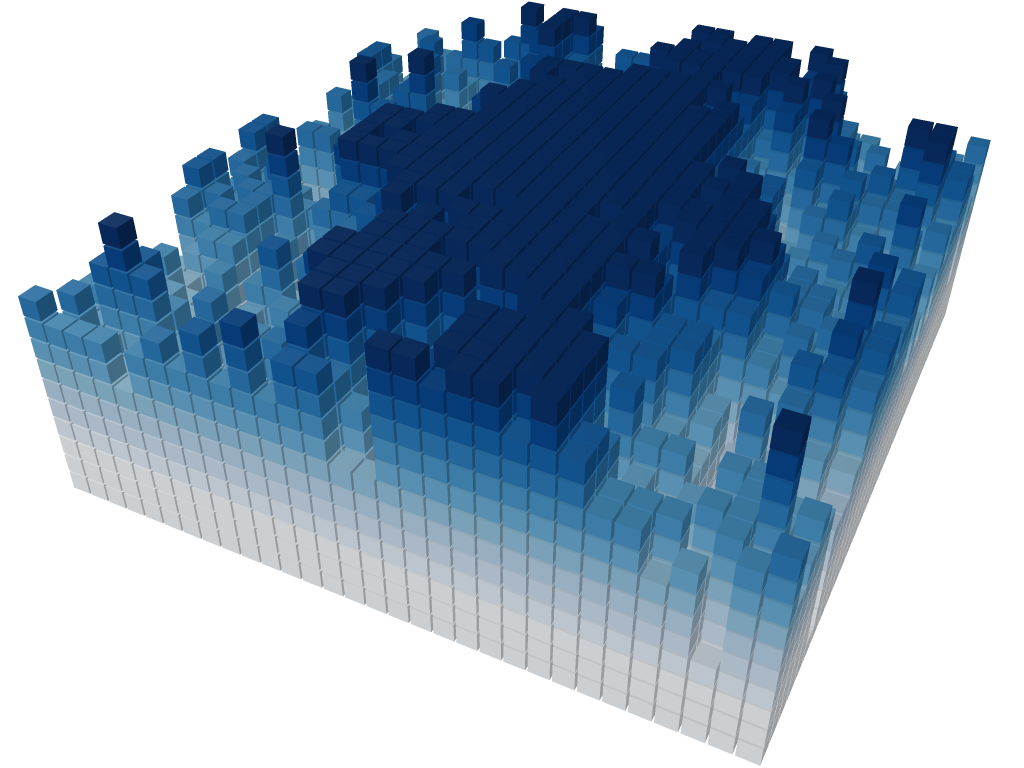} & \hspace{\colsepeqs} & 
  \includegraphics[width=\eqswtwo\textwidth]{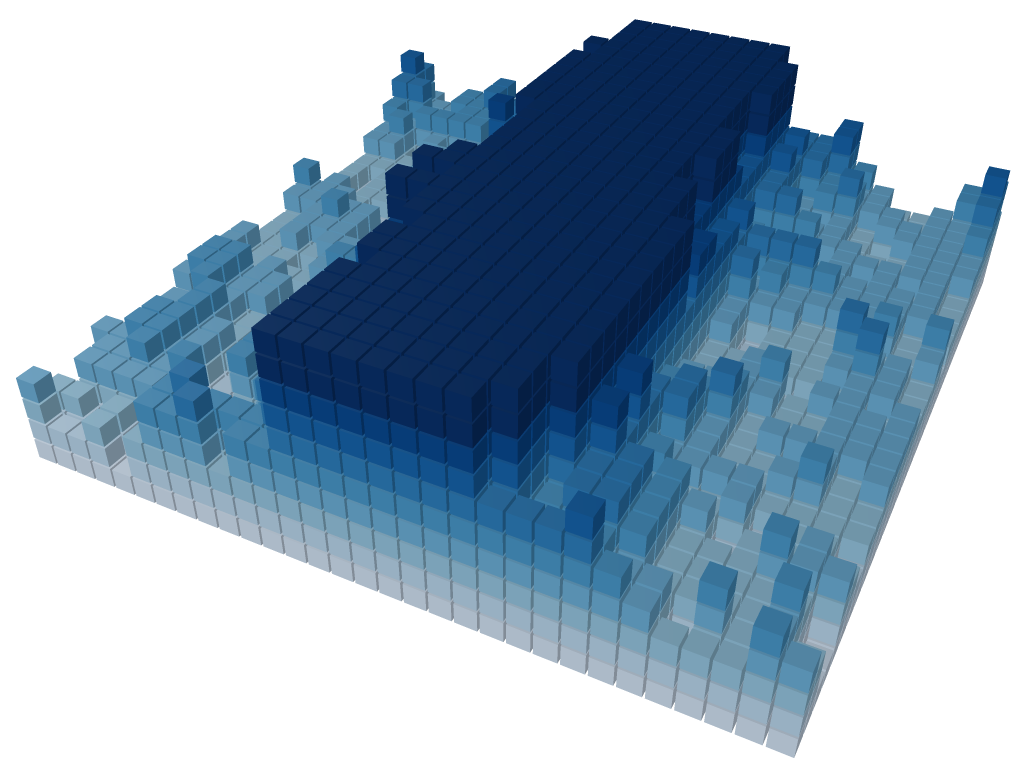} & \hspace{\colsepeqs} & 
  \includegraphics[width=\eqswtwo\textwidth]{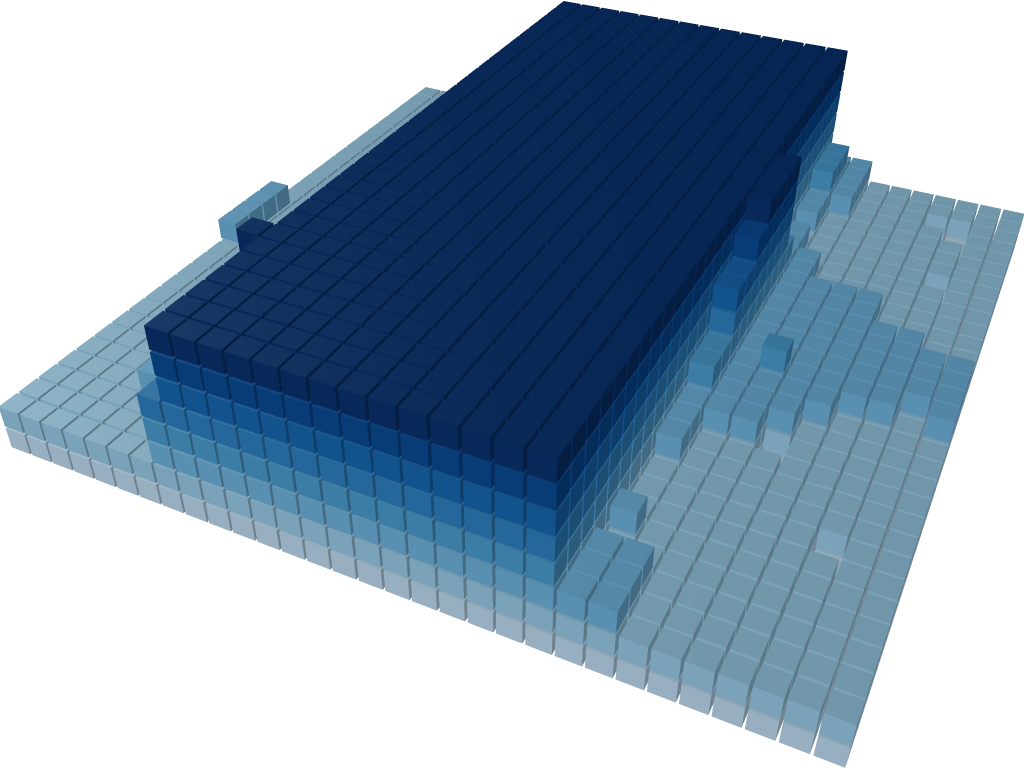} \\
  $F_0/E_bA = \ifzii$ \hspace{1cm} & \hspace{\colsepeqs} & $F_0/E_bA = \iifzii$ \hspace{1cm} & \hspace{\colsepeqs} & $F_0/E_bA = \iiifzii$ \hspace{1cm} \\
  $\rhoWV = \irii$    \hspace{1cm} & \hspace{\colsepeqs} & $\rhoWV = \iirii$    \hspace{1cm} & \hspace{\colsepeqs} & $\rhoWV = \iiirii$    \hspace{1cm} \\
  & & & & \\
  \includegraphics[width=\eqswtwo\textwidth]{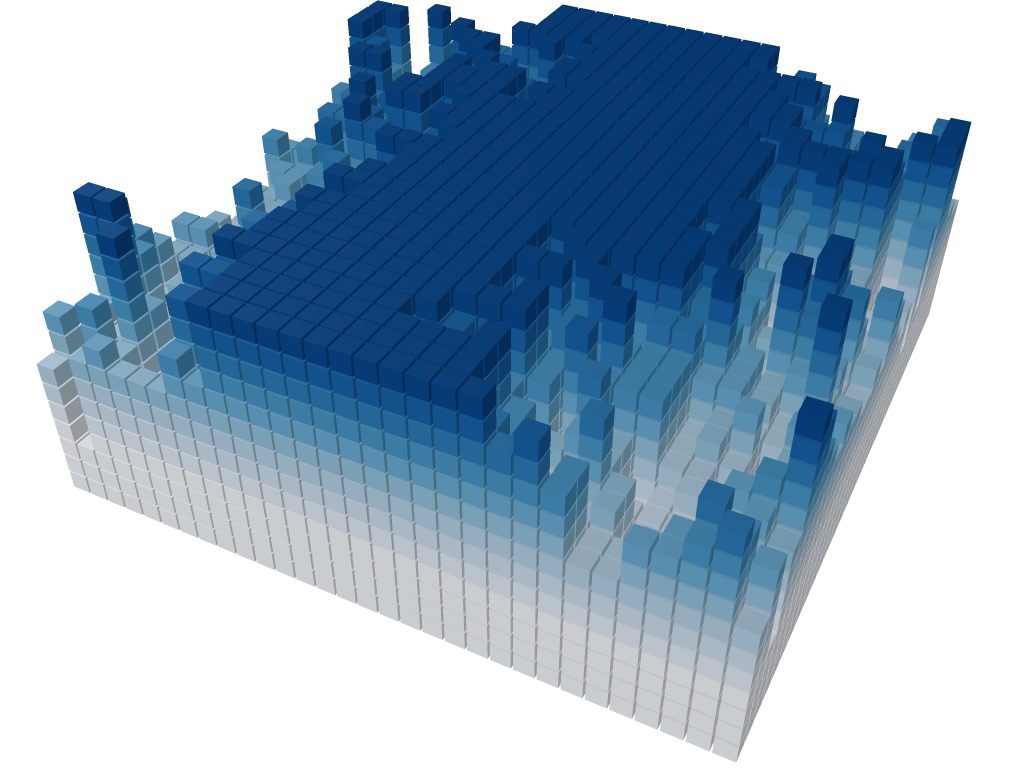} & \hspace{\colsepeqs} & 
  \includegraphics[width=\eqswtwo\textwidth]{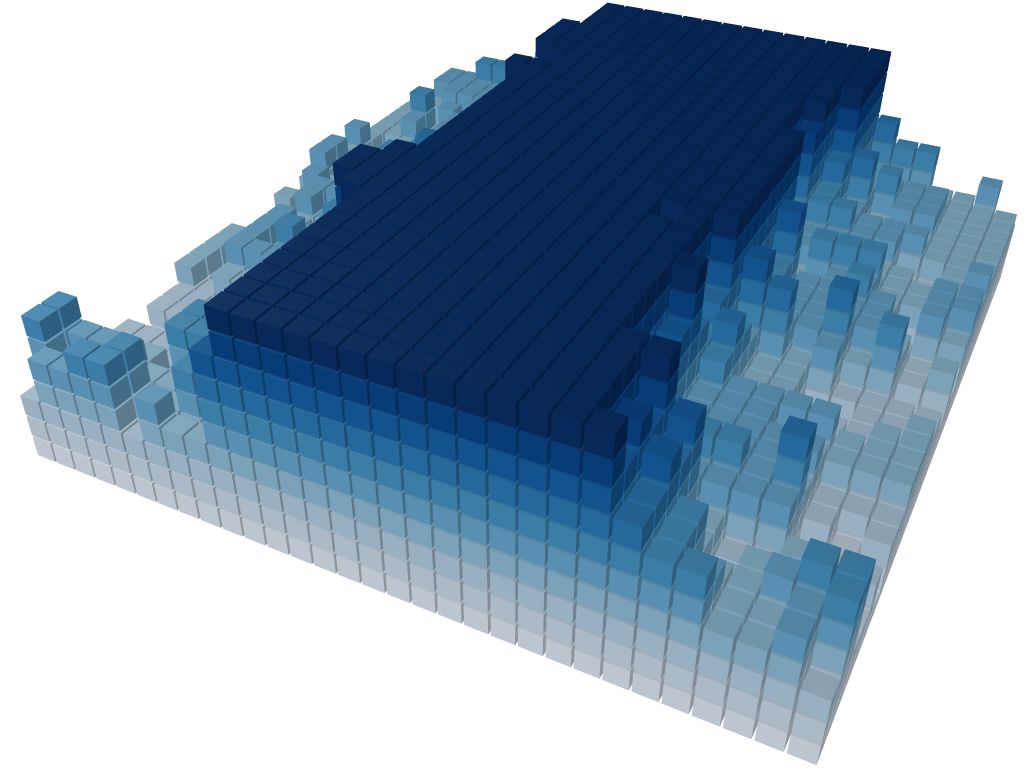} & \hspace{\colsepeqs} & 
  \includegraphics[width=\eqswtwo\textwidth]{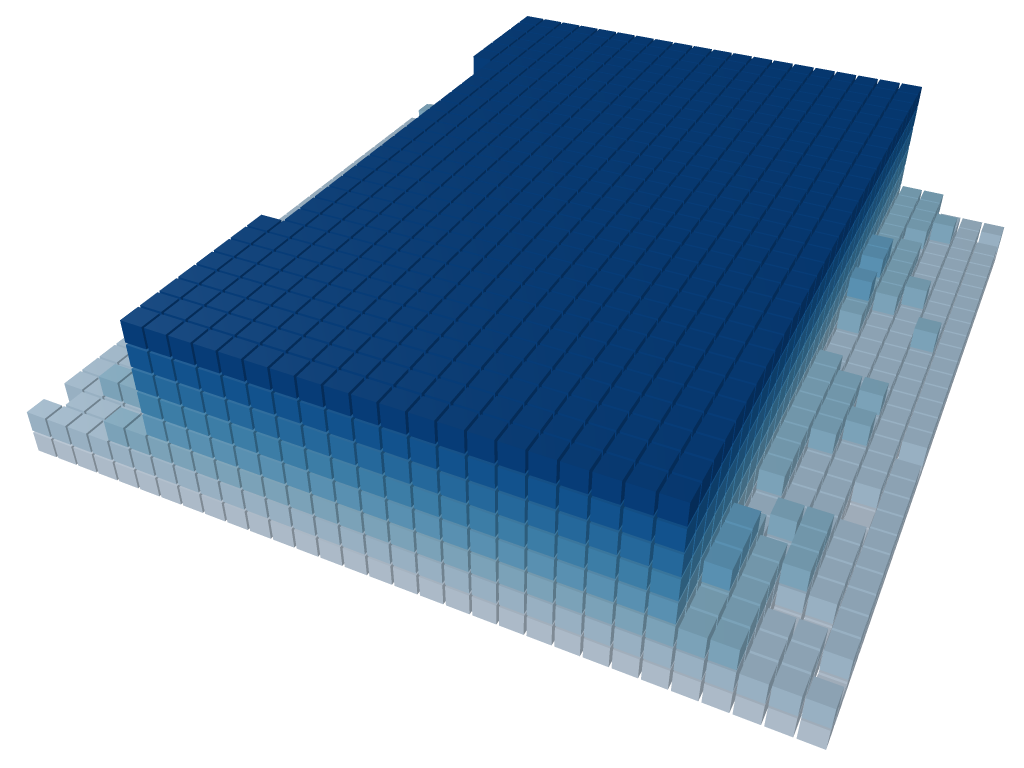} \\
  $F_0/E_bA = \ifziii$ \hspace{1cm} & \hspace{\colsepeqs} & $F_0/E_bA = \iifziii$ \hspace{1cm} & \hspace{\colsepeqs} & $F_0/E_bA = \iiifziii$ \hspace{1cm} \\
  $\rhoWV = \iriii$    \hspace{1cm} & \hspace{\colsepeqs} & $\rhoWV = \iiriii$    \hspace{1cm} & \hspace{\colsepeqs} & $\rhoWV = \iiiriii$    \hspace{1cm} \\
  & & & & \\
  \includegraphics[width=\eqswtwo\textwidth]{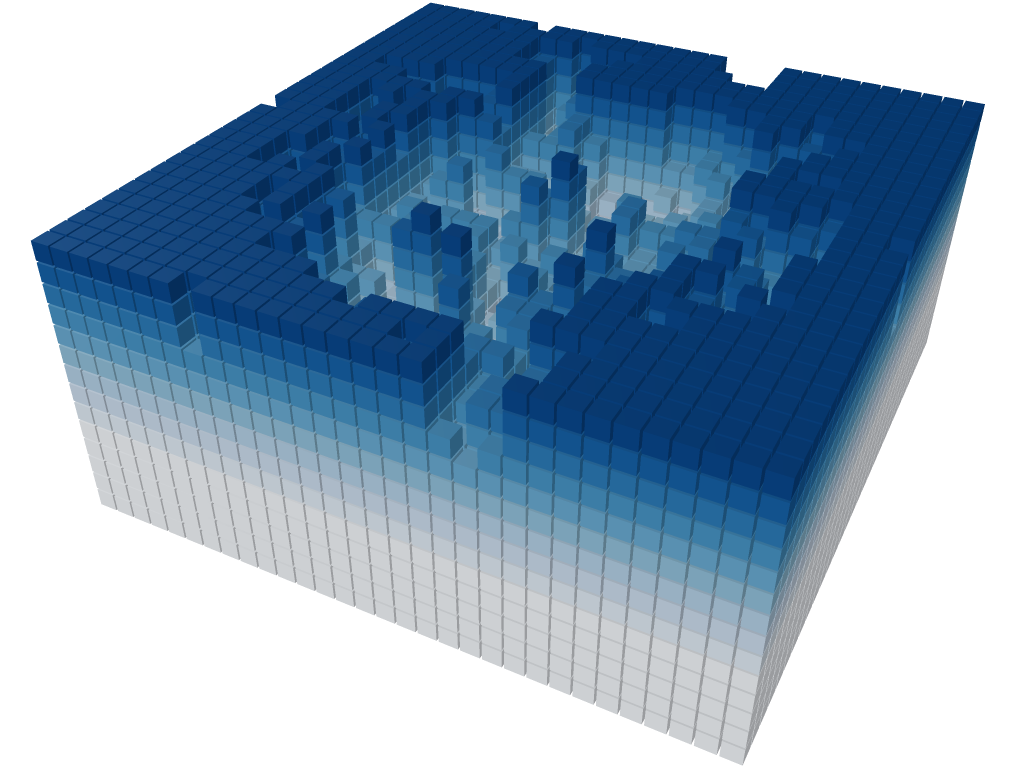} & \hspace{\colsepeqs} & 
  \includegraphics[width=\eqswtwo\textwidth]{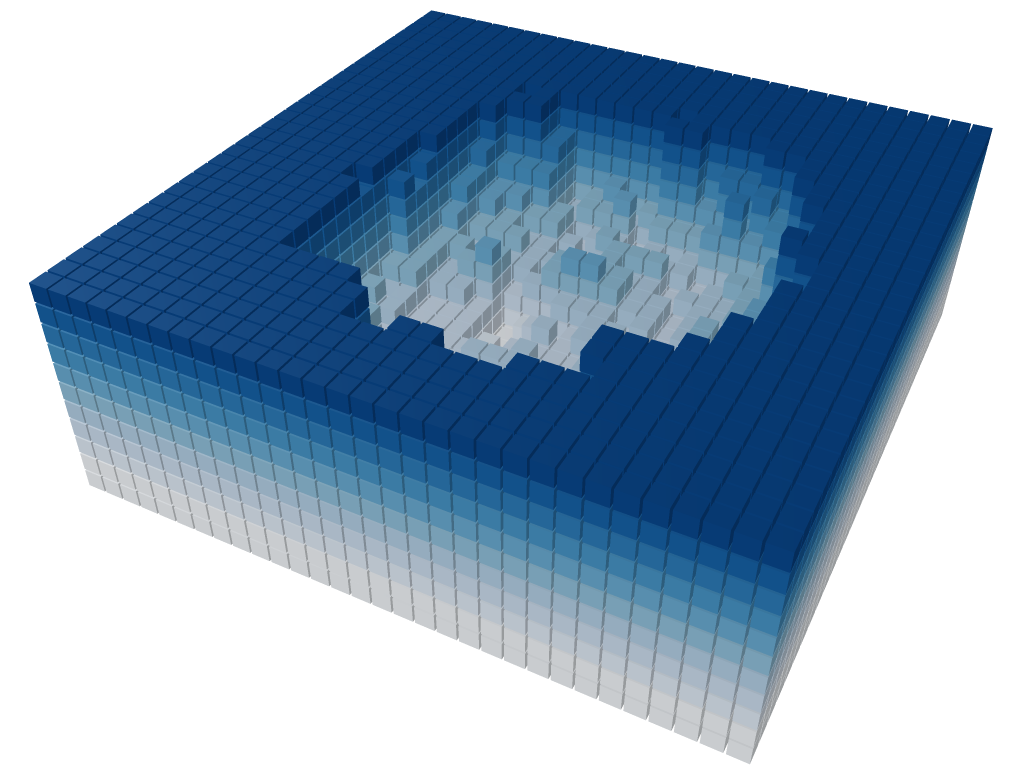} & \hspace{\colsepeqs} & 
  \includegraphics[width=\eqswtwo\textwidth]{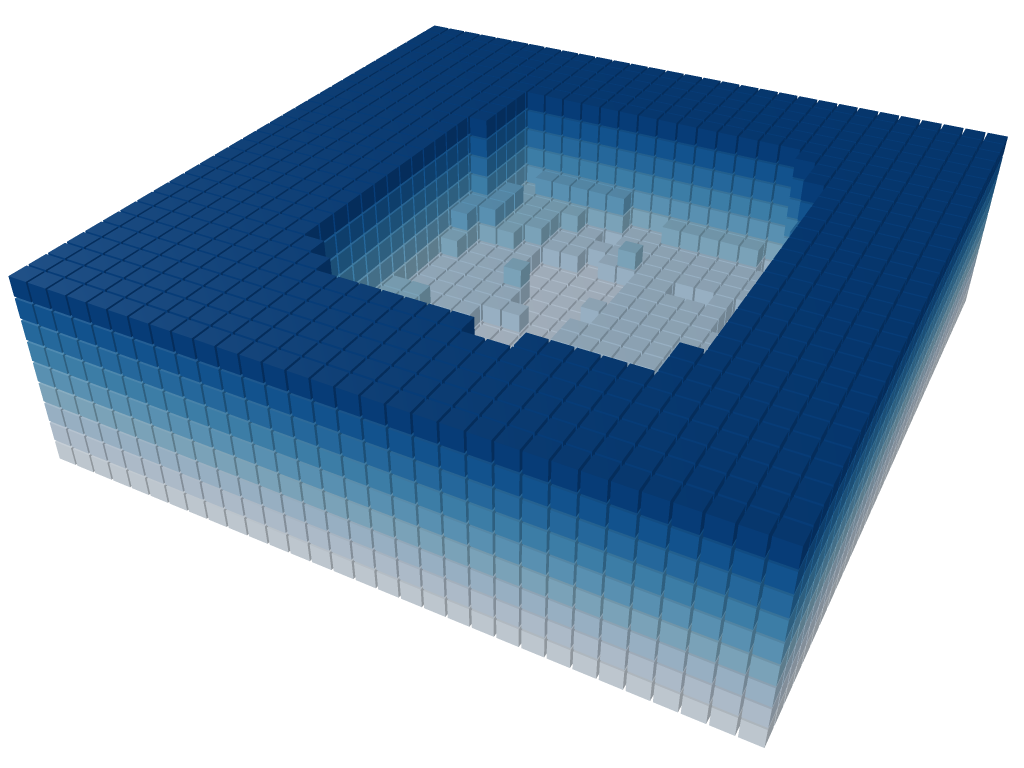} \\
  $F_0/E_bA = \ifziv$ \hspace{1cm} & \hspace{\colsepeqs} & $F_0/E_bA = \iifziv$ \hspace{1cm} & \hspace{\colsepeqs} & $F_0/E_bA = \iiifziv$ \hspace{1cm} \\
  $\rhoWV = \iriv$    \hspace{1cm} & \hspace{\colsepeqs} & $\rhoWV = \iiriv$    \hspace{1cm} & \hspace{\colsepeqs} & $\rhoWV = \iiiriv$    \hspace{1cm} \\
  & & & & \\
  \includegraphics[width=\eqswtwo\textwidth]{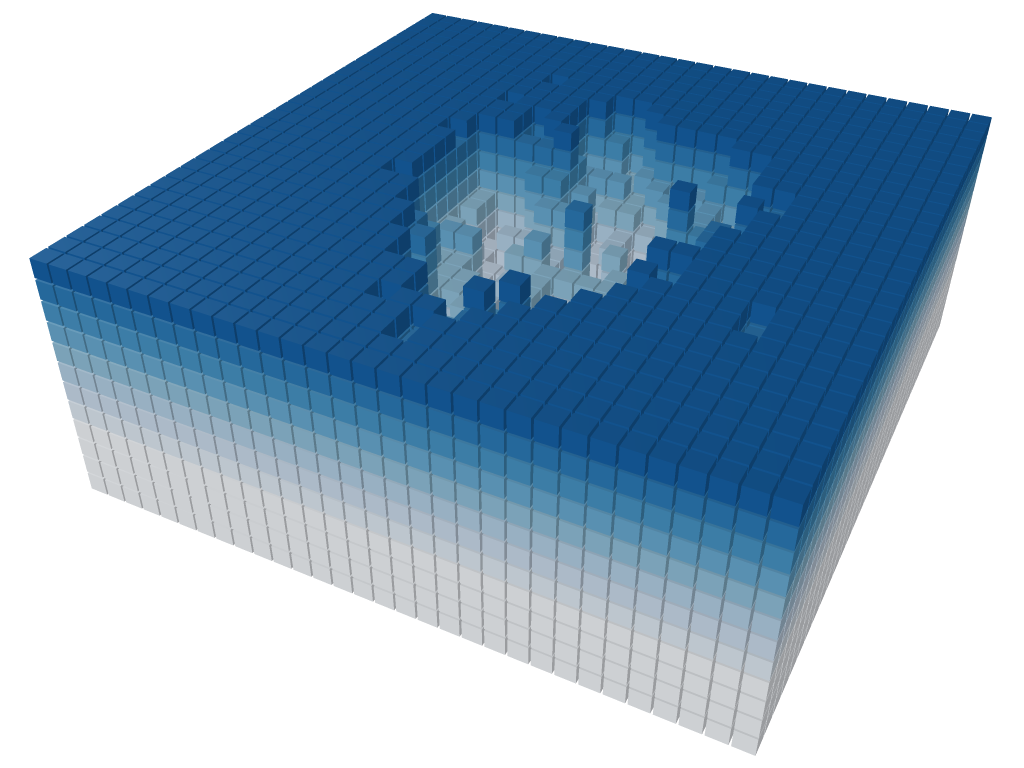} & \hspace{\colsepeqs} & 
  \includegraphics[width=\eqswtwo\textwidth]{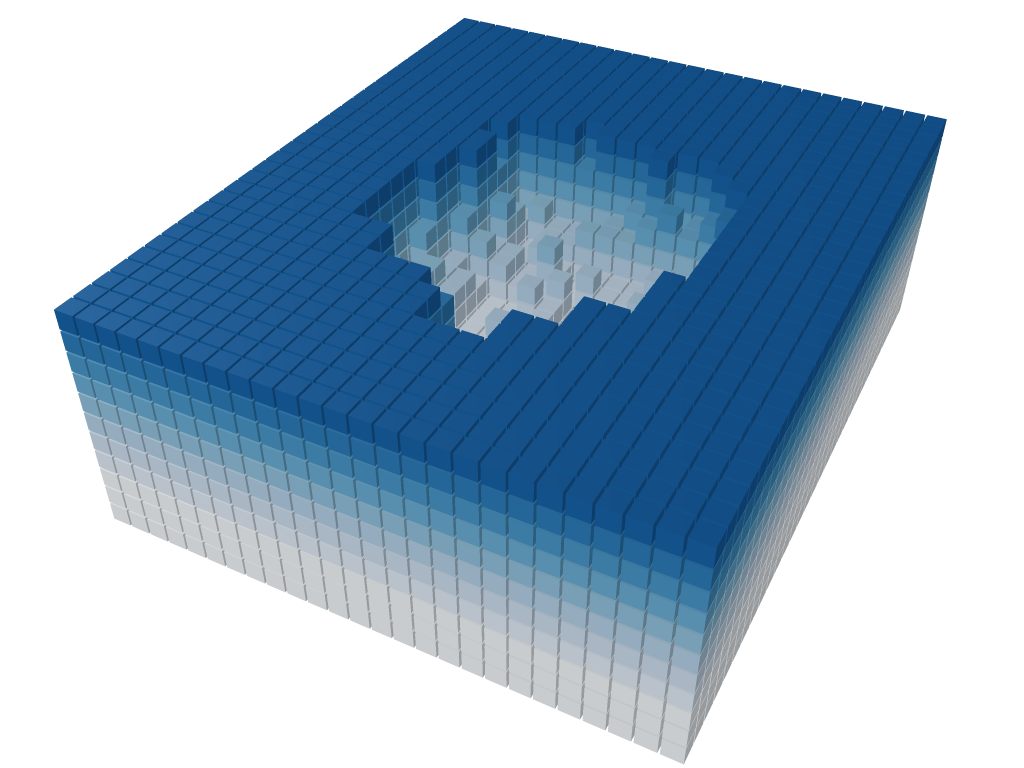} & \hspace{\colsepeqs} & 
  \includegraphics[width=\eqswtwo\textwidth]{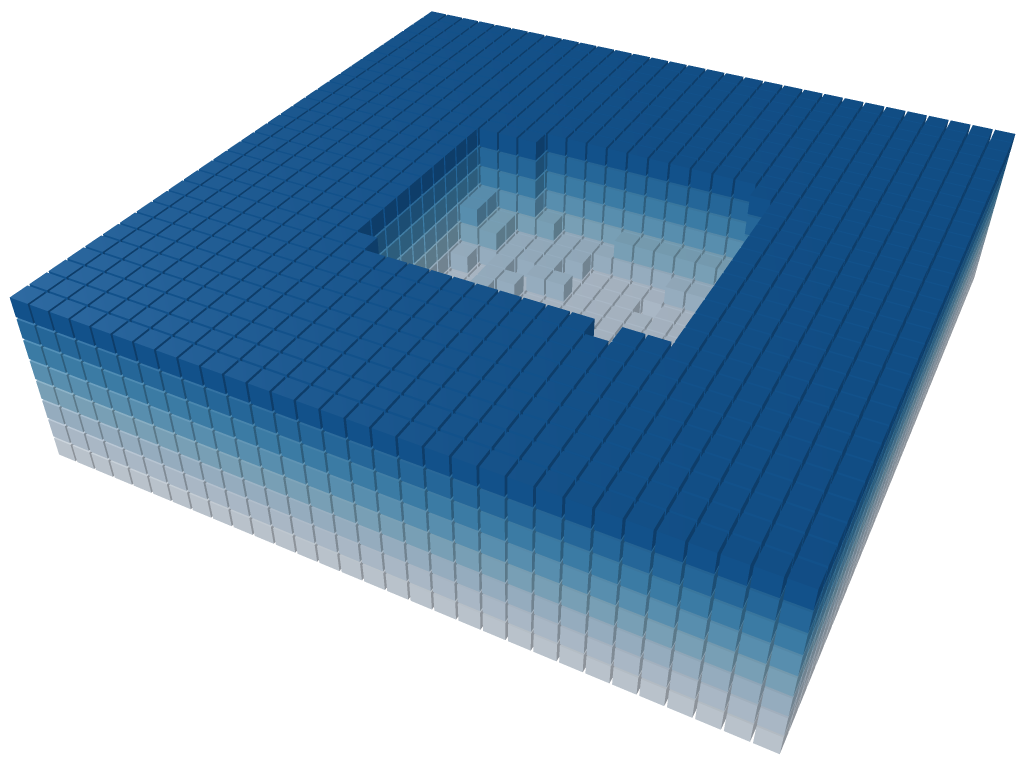} \\
  $F_0/E_bA = \ifzv$ \hspace{1cm} & \hspace{\colsepeqs} & $F_0/E_bA = \iifzv$ \hspace{1cm} & \hspace{\colsepeqs} & $F_0/E_bA = \iiifzv$ \hspace{1cm} \\
  $\rhoWV = \irv$    \hspace{1cm} & \hspace{\colsepeqs} & $\rhoWV = \iirv$    \hspace{1cm} & \hspace{\colsepeqs} & $\rhoWV = \iiirv$    \hspace{1cm} \\
  \hline \\
 \end{tabular}
  \caption{Equilibrium shapes with a given contact density $\rhoWV$ at rescaled inverse temperature $E_b/kT$ and rescaled external pressure $F_0/E_bA$ using $\sigma_0=1$. 
  We observe three classes of shapes: islands, bands and pits. The images have been rendered using Ovito~\cite{Stukowski2009a}.}
  \label{fig:eq_shapes}
 \end{center}
\end{figure*}

In this section we investigate what determines the equilibrium contact size,
shape and stability.

In Fig.~\ref{fig:eq_shapes} we show a selection of equilibrium shapes obtained by simulations
at different rescaled temperatures $kT/E_b$, rescaled applied pressures $F_0/E_bA$ using $\sigma_0 = 1$.
We see three classes of shapes occurring: islands for low contact densities, periodic bands 
for intermediate contact densities, and pits for large contact densities.
The shapes appear rougher for higher temperatures, which is expected due to 
larger surface fluctuations.

It makes sense that the resulting shape should depend on the contact density, 
since we can imagine an island transforming into a band when it coalesces with its own
periodic image, and the pits forming when the bands grow so wide they merge with their own periodic images.

In practice surface dislocations and/or solute diffusion could set a characteristic length scale 
promoting the existence of separate islands. These islands 
could merge by forming bands between one another, 
which in several directions would lead to pit shapes.
Nucleation on system boundaries could also lead to half-islands, 
bands and pits depending on the size of the crystals growing on the boundaries.

It is clear that the equilibrium contact density is a key factor in deciding the behavior 
of the equilibrium contact. Since the contact density is not a priori known,
we need to investigate how the contact density depends on the system parameters.
This relationship is shown in Fig.~\ref{fig:model_results}.

For each value of $\sigma_0$ we observe a domain where $\rhoWV = 0$, i.e. no stable contacts are formed. 
This domain represents the cases where the surface fluctuations are not large enough relative to the surface-surface separation 
for stable contacts to form. Since applying more pressure 
decreases the separation,
and increasing the temperature increases the surface fluctuations, 
we see this domain curving off with higher pressure $F_0/E_bA$ and higher $E_b/kT$.

At higher pressures, contact initiation is not limited by surface fluctuations, since the initial separation is low.
In this limit, higher fluctuations simply means that the surfaces can fit less tight.
Hence for large pressures we observe that increasing $E_b/kT$ results in larger contacts.

\newcommand{\fullfigheight}{6.5cm}
\begin{figure*}[ht]
 \begin{center}
  \begin{tabular}{ccc} 
   \includegraphics[height=\fullfigheight]{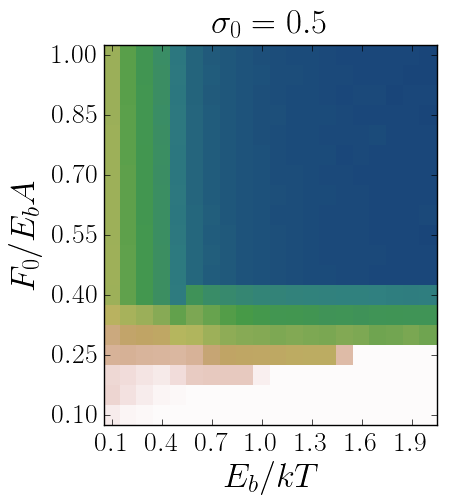} &
   \includegraphics[height=\fullfigheight]{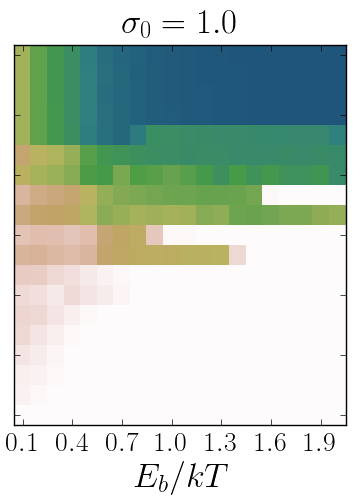} &
   \includegraphics[height=\fullfigheight]{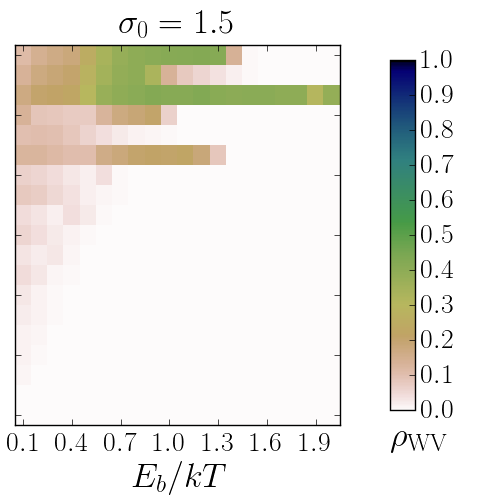} \\
  \end{tabular}
 \end{center}
 \caption{Plot of the contact density $\rhoWV$ vs rescaled inverse temperature $E_b/kT$ and rescaled external pressure $F_0/E_bA$
 for various repulsion strengths $\sigma_0$. The latter is relative to the bond energy $E_b$. Each data point is averaged over ten simulations.
 A white field means no stable contacts were formed. 
 Higher pressures generally means more contacts since the surfaces are closer, and thus we see more contacts at the top of each plot than at the bottom.
 The reverse occurs when we increase $\sigma_0$ (more repulsion), which is why the right figure has a lot less contacts than the left.
 As we heat the system by lowering $E_b/kT$, we observe fewer contacts, 
 which is due to the surface fluctuations becoming very high, which destabilizes contacts. Hence we see a drop in $\rhoWV$ 
 as we move left in each plot.
 The surface height fluctuations are those responsible for creating the initial stable contacts,
 hence cooling the system by increasing $E_b/kT$ makes it very hard to initiate contacts and we see a drop in $\rhoWV$ after a certain point,
 unless the pressures is so high that contacts are inevitable. If contacts are inevitable, 
 then lowering the fluctuations means that the surfaces can fit more smoothly on top of one another, and we expect an increase in $\rhoWV$.
 Hence we observe a decay in $\rhoWV$ to the right in each plot for low pressure and an increase in $\rhoWV$ to the right in each plot for high pressures. 
 For higher $\sigma_0$, we see lines at certain pressures that decays slower
 with increasing $E_b/kT$ than their surroundings, hinting to a mechanism that favors contacts to form at certain pressures.}
\label{fig:model_results}
\end{figure*}
\begin{figure*}
 \begin{center}
  \begin{tabular}{ccc}
   \includegraphics[height=\fullfigheight]{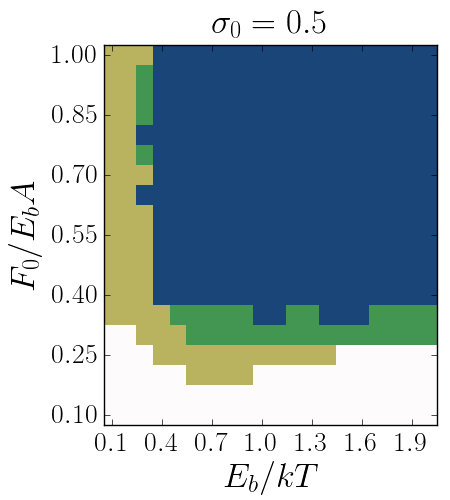} &
   \includegraphics[height=\fullfigheight]{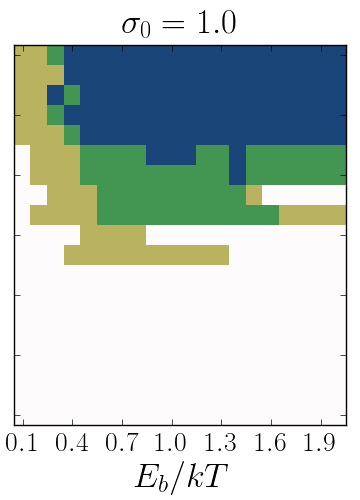} &
   \includegraphics[height=\fullfigheight]{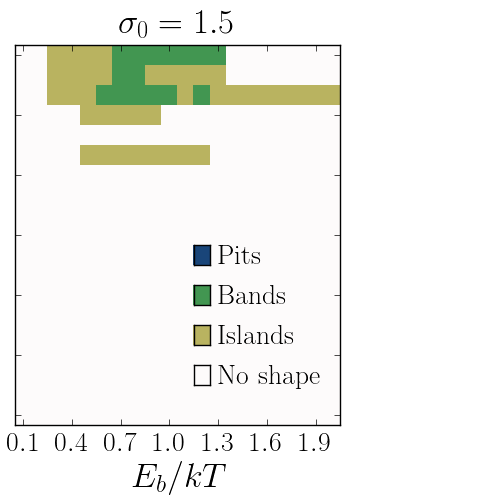} \\
  \end{tabular}
 \end{center}
 \caption{Plot of the most frequently occurring equilibrium contact cluster shape for a given rescaled external pressure $F_0/E_bA$, rescaled inverse temperature $E_b/kT$ and repulsion strength $\sigma_0$. 
 The latter is relative to the bond energy $E_b$. 
 These data are obtained by pattern recognition of the same surfaces used to produce Fig.~\ref{fig:model_results}. 
 It is clear that there is a strong correlation between the contact density $\rhoWV$ and the stability of the different shapes.
 The parameter dependency explained in Fig.~\ref{fig:model_results} thus holds for this figure as well,
 with the exception that some low contact density cases are missing since they do not correspond to any shape.
 }
\label{fig:phasediag}
\end{figure*}

\subsubsection{Contact favoring pressure levels}
\label{sec:resonance}

At high pressures the surface-surface separation is so low that contact clusters always form. 
An interesting effect which we will investigate in this section is that 
at certain lower pressures contact clusters appear much more stable than their immediately lower and higher pressure levels.
This can be seen as stripes going into the unstable domain ($\rhoWV = 0$) in Fig.~\ref{fig:model_results}.

A contact favoring pressure level (CFPL) appears because the corresponding far equilibrium point of the confining surface $h_\lambda$
is very close to an integer value.
This means that when the crystal surface fluctuates into contact with the confining surface,
the binding energy is close to the maximum value $E_b$ at a separation of $d_i=1$, 
which makes it harder to remove than if the separation was anything else.

For simplicity lets consider a flat crystal surface at an integer height $\tilde h$.
We know from Eq.~(\ref{eq:h_lambda_analytical}) that the far equilibrium point is

\begin{equation}
 \tilde h_\lambda = \tilde h + 1 + \lambda_D \ln \left(\frac{\sigma_0/\lambda_D\xi }{F_0/E_bA}\right),
\end{equation}

\noindent 
if there are no attractive interactions.

The condition that this height is an integer value $n$ then translates into the following equation:

\begin{equation}
\label{eq:resonance_criteria}
 \lambda_D \ln \left(\frac{\sigma_0/\lambda_D\xi }{F_0/E_bA}\right) = n,
\end{equation}

\noindent
The CFPLs then become

\begin{equation}
 F_n/E_bA = \frac{\sigma_0}{\lambda_D\xi}\exp(-n/\lambda_D),
\end{equation}

\noindent
and the distance between two sequential CFPLs becomes

\begin{align}
 \Delta P_n &= \frac{\sigma_0}{\lambda_D\xi}\left[\exp(-n/\lambda_D) - \exp(-(n+1)/\lambda_D\right] \notag \\
            &= \frac{\sigma_0}{\lambda_D}\exp(-n/\lambda_D).
\end{align}

\noindent
These equations are in agreement with the fact that the CFPLs seems further apart for larger $\sigma_0$.

\begin{figure}[ht]
 \begin{center}
  \includegraphics[width=1.0\columnwidth]{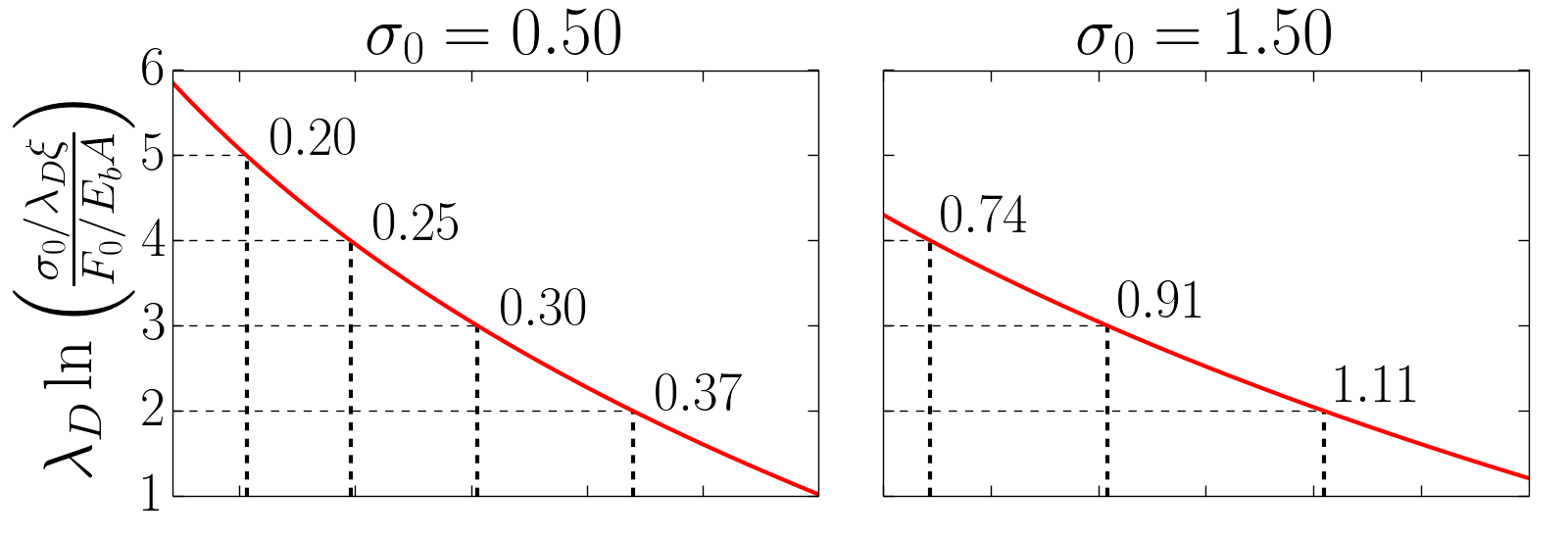}
  \includegraphics[width=1.0\columnwidth]{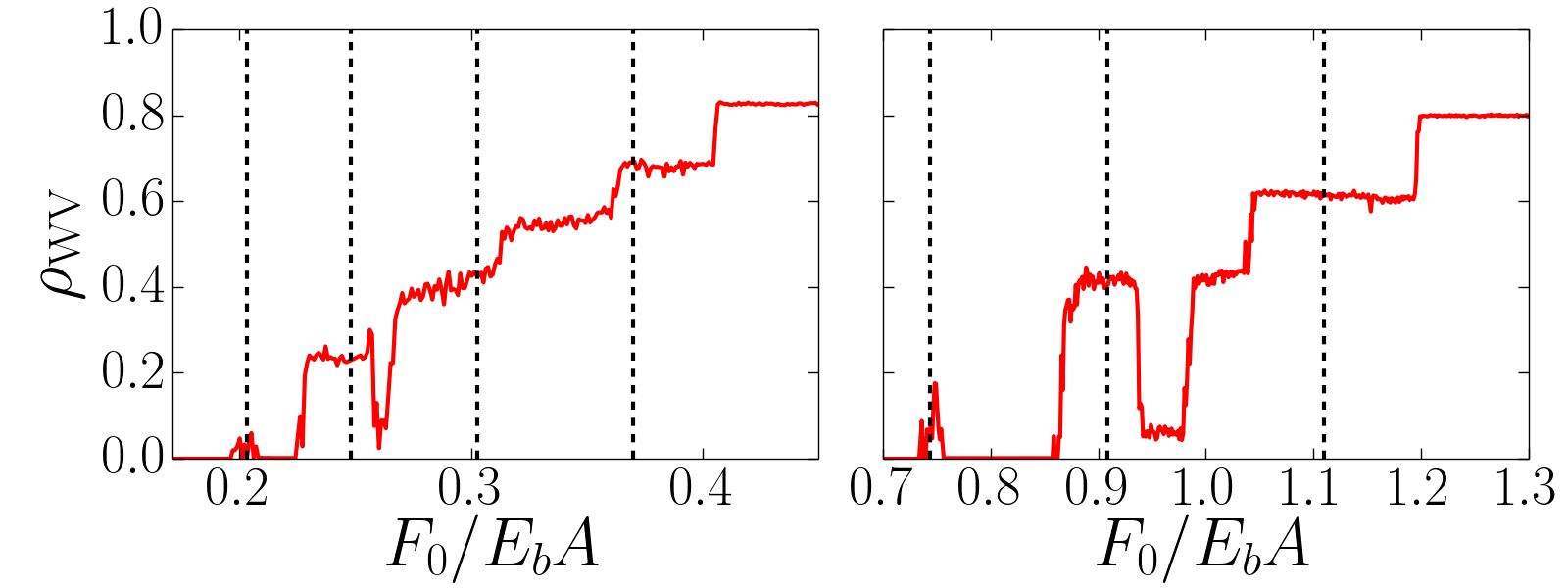}
  \caption{Plot of the critera in Eq.~(\ref{eq:resonance_criteria}) (top row) and contact density $\rhoWV$ (bottom row) vs rescaled external pressure $F_0/E_bA$ for 
   $\sigma_0=0.5$ (left column) and $\sigma_0=1.5$ (right column). 
   The simulations in the bottom row are performed using $E_b/kT = 1.2$.
  The top row predicts contact favoring pressure levels (CFPLs) when the function takes integer values. The stippled lines are visual aids
  to identify these levels. We see that Eq.~(\ref{eq:resonance_criteria}) matches very well where the levels are sparse, 
  but when they become crowded we miss out on some. This is expected since when the contact density increases, the attractive interactions, 
  which we completely ignored when deriving Eq.~(\ref{eq:resonance_criteria}), becomes increasingly important.}
 \label{fig:simulated_resonance}
 \end{center}
\end{figure}

In Fig.~\ref{fig:simulated_resonance} we show the predicted CFPLs (top row) together with simulations 
done at a specific value of $E_b/kT=1.2$ (bottom row). It is clear that there analytical predictions matches very well
with the simulations, however, some levels are not predicted. This is expected since 
we in the derivation completely ignored the attractive interactions.

The reason why the CFPLs appear flat is that if there is a small gap, 
the surfaces will snap into contact, resulting in the same scenario as if the confining surface was located at an integer value. 
This also explains why the contact density appears to possess discrete levels and transitions from one level to another quite abruptly.	

If we instead of solving a mechanical equilibrium problem simply fixed the position of the confining surface,
we would expect the contact density to oscillate smoothly as the confining surface position was varied.
This behavior is phenomenologically similar to oscillatory hydration forces~\cite{Israelachvili2011i} 
where the energy cost of removing layers of water confined between surfaces oscillates with a period related to the thickness of the layers.

\subsubsection{Equilibrium contact shape vs size}
\label{sec:eq_shape_vs_size}

The contact density $\rhoWV$ does not contain any information about the shape of the contact. 
However, by comparing the equilibrium densities in Fig.~\ref{fig:model_results} to the most frequently occurring equilibrium shapes in Fig.~\ref{fig:phasediag},
we see that the two are strongly correlated.

The equilibrium contact shape minimizes the free energy of the contact, and for high values of $E_b/kT$
the free energy is dominated by the binding energies.

The free energy due to binding energies is given by the number of available bonds into solution~\cite{Hogberget2016},
which means we expect square-like contact shapes for a cubic lattice structure. 
For lower $E_b/kT$ we would expect the shapes to fluctuate around these square-like shapes.

For simplicity we will consider only square surfaces with area $L^2$ and square contact shapes with area $A_c = L^2\rhoWV$.
For a square island shape, the relationship between the contact area and the 
contact perimeter $S_I$ is

\begin{equation}
 S_I = 4\sqrt{A_c} = 4L \sqrt{\rhoWV}.
\end{equation}

For the bands we have no perimeter associated with the width of the contact, 
hence the perimeter is simply

\begin{equation}
 S_B = 2L.
\end{equation}

The pit shapes are inverted versions of the islands.
The area of the pit cavity is $\bar{A} = L^2 - A_c = L^2(1 - \rhoWV)$, such that the perimeter becomes

\begin{equation}
 S_P = 4\sqrt{\bar{A}} = 4L\sqrt{1 - \rhoWV}.
\end{equation}

In order to predict which of these shapes are the most stable ones, 
we need to know which of these three cases has the shortest perimeter
for a given value of $\rhoWV$. Comparing $S_I$ and $S_B$ yields

\newcommand{\RHOSEP}{\quad\rightarrow\quad}
\begin{equation}
\label{eq:rho_lim_lower}
 S_I < S_B \RHOSEP \rhoWV < 1/4,
\end{equation}

\noindent
which means that for contact densities lower than $1/4$, we expect islands. 
Comparing $S_B$ and $S_P$ yields

\begin{equation}
\label{eq:rho_lim_upper}
 S_P < S_B \RHOSEP 1 - \rhoWV < 1/4 \RHOSEP \rhoWV > 3/4,
\end{equation}

\noindent
which means that for contact densities higher than $3/4$, we expect pits.

\begin{figure}
 \begin{center}
  \includegraphics[width=0.4\paperwidth]{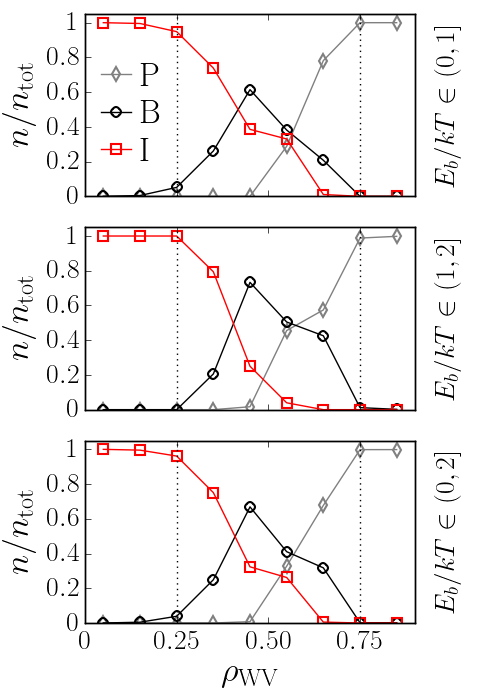}
  \caption{Plot of the number fraction of occurring phases $n/n_\mathrm{tot}$ as a function of the contact density level $\rhoWV$
  for three temperature intervals: high temperatures (top panel), low temperatures (middle panel),
  and all temperatures (bottom panel). For small contacts we get only islands (I), for medium contacts where $\rhoWV\in [1/4, 3/4]$ 
  we get bands (B) as well, and above this limit we get mostly pits (P). 
  These results are obtained by combining Fig.~\ref{fig:model_results} with Fig.~\ref{fig:phasediag}. 
  The dashed lines are the theoretical stability limits for the bands from Eqs.~(\ref{eq:rho_lim_lower}) and (\ref{eq:rho_lim_upper}).
  It is clear that they agree well with the simulations.}
  \label{fig:prop_cov}
 \end{center}
\end{figure}

Equivalently we may say that we expect bands when $\rhoWV\in [1/4, 3/4]$.
In Fig.~\ref{fig:prop_cov} we show the number density of an occurring shape
as a function of $\rhoWV$, and we see that the limits derived here agree with the simulations. 
We also see that islands and pits cannot be stable at the same value of $\rhoWV$ without bands also being stable,
which means that the bands are a transitional state between islands and pits.

\newcommand{\fflucfigh}{7.25cm}
\newcommand{\flucfigh}{8.5cm}
\begin{figure*}
 \begin{center}
  \includegraphics[height=\fflucfigh]{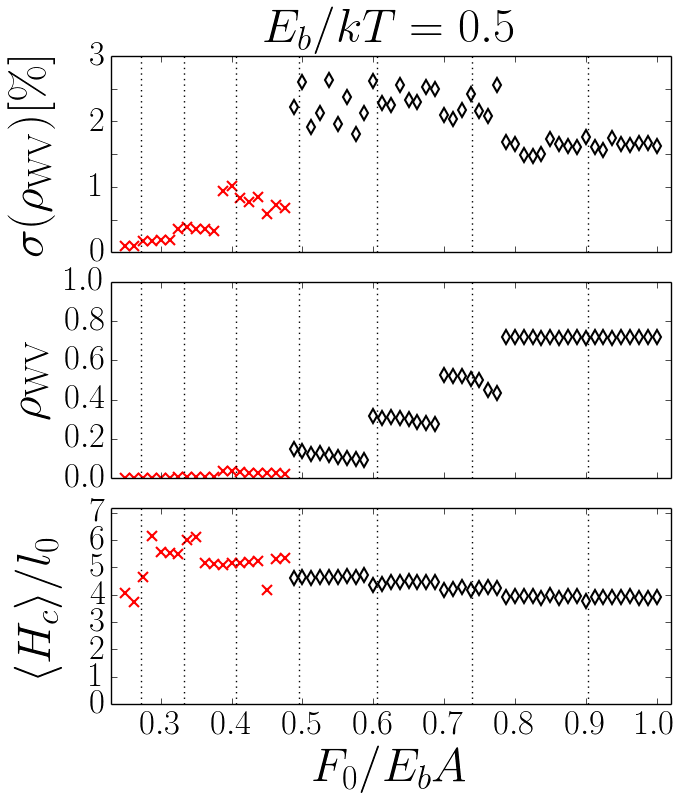} 
  \includegraphics[height=\fflucfigh]{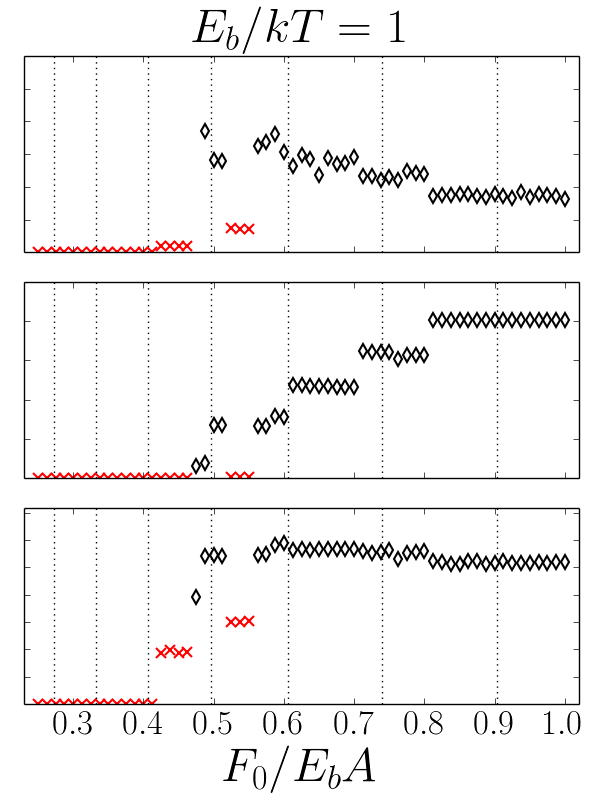}
  \includegraphics[height=\fflucfigh]{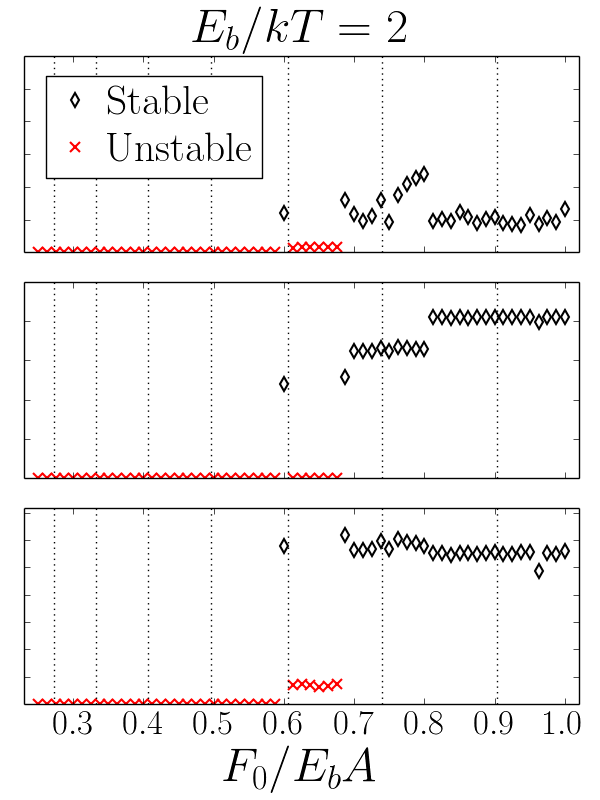} \\
  \caption{The contact density $\rhoWV$ and its fluctuations $\sigma(\rhoWV)$, measured as the standard deviation 
  of the time series of $\rhoWV$ when the contacts have stabilized, vs applied pressure $F_0/E_bA$ using $\sigma_0 = 1$.
  The vertical stippled lines are the contact favoring pressure levels (CFPLs) given by Eq.~(\ref{eq:resonance_criteria}). 
  The bottom row shows the average contact clusters height $\langle \Delta H_c\rangle/l_0$ defined in Eq.~(\ref{eq:contact_height}).
  The crosses represent simulations where no stable contact clusters were formed. 
  Since the thermal fluctuations of the surface increase with temperature,
  and the contact is part of the surface,
  the contact fluctuations also follow this trend. We also see that the contact cluster fluctuations decrease as $\rhoWV$ increases. 
  The jumps in the fluctuations are clearly caused by $\rhoWV$ transitioning to a different CFPL.}
  \label{fig:fluctuations_rho}
 \end{center}
 \end{figure*}
\begin{figure*}
 \begin{center}
  \includegraphics[height=\flucfigh]{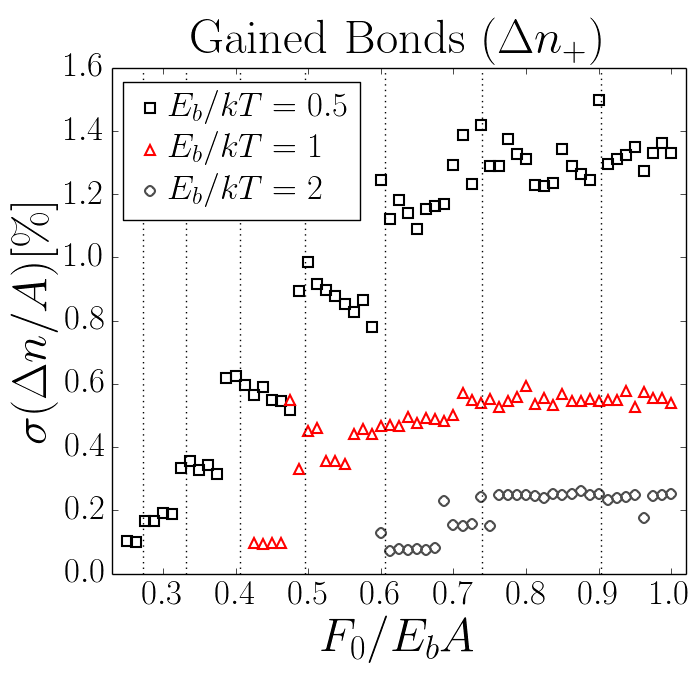} 
  \includegraphics[height=\flucfigh]{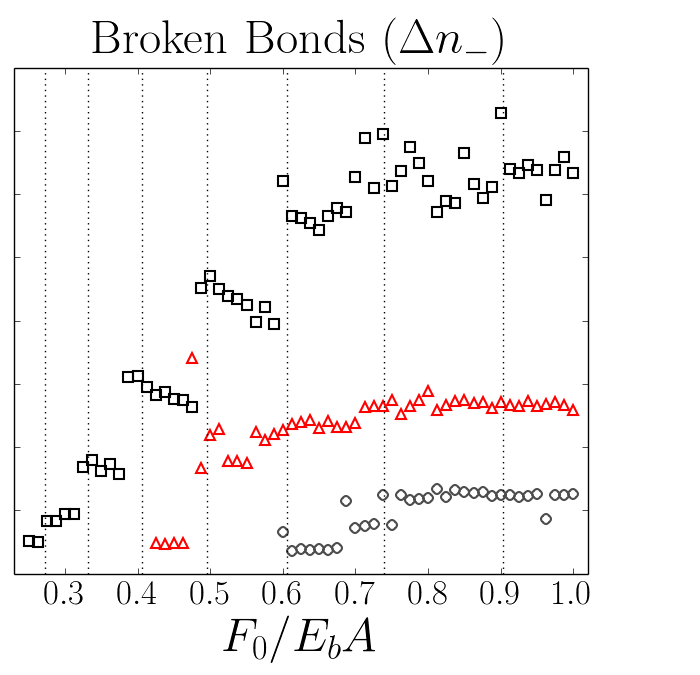} \\
  \caption{Fluctuations in the number of broken bonds $\Delta n_-$ and number of gained bonds $\Delta n_+$ sampled every 10 000 time steps
  for various rescaled inverse temperatures $E_b/kT$ and rescaled external pressures $F_0/E_bA$ using $\sigma_0 = 1$. The vertical dashed lines represent the contact favoring pressure levels (CFPLs)
  calculated by Eq.~(\ref{eq:resonance_criteria}). As expected, increasing the temperature increases the fluctuations. 
  The fluctuations stagnate at the point where the surfaces are initiated in contact, such that applying more pressure has no effect.
  We clearly see a correlation between the jumps in fluctuations and the CFPLs. }
  \label{fig:fluctuations_n}
 \end{center}
\end{figure*}

\subsection{Contact fluctuations}
\label{sec:fluctuations}

In this section we investigate how a stable contact fluctuates in time. 
In e.g.~Fig.~\ref{fig:equilibation} we observed that the contact clusters were dynamic,
and the contact density $\rhoWV$ fluctuated around a mean value in time.
We measure these fluctuations as the standard deviation of the time series of $\rhoWV$ in equilibrium, 
that is

\begin{equation}
 \sigma(\rhoWV) = \sqrt{\langle (\rhoWV - \langle \rhoWV \rangle_t)^2\rangle_t},
\end{equation}

\noindent
where $\langle X\rangle_t$ denotes the average of $X$ in time.

In the top and middle row of Fig.~\ref{fig:fluctuations_rho} these fluctuations and the corresponding 
value of $\rhoWV$ are presented for various pressures. We see that the fluctuations 
increase with increasing temperature while the dependency on the pressure decreases.
This happens since for high temperature systems, thermal surface fluctuations are dominant.
At high pressures the surfaces are always resting on one another,
which means that applying more pressure has no effect. Hence 
the fluctuations stagnate.

Except for high temperature systems ($E_b/kT=0.5$), 
we see that larger contact clusters have lower  
fluctuations. 
This is to be expected since the stability of a contact depends on its total number of nearest neighbor bonds,
and large compact clusters have a low surface to area ratio.
Hence these results suggest that the mobility of a contact cluster increases with increasing temperature and decreasing size.

Fluctuations in the contact cluster size and the surface roughness are correlated since 
the more the surface heights are fluctuating, the more often we expect the surface to transition in and out of the contact regime.
This is the same mechanism that correlated the pressure (initial height) and the contact size $\rhoWV$
in Fig.~\ref{fig:model_results}, i.e.~smaller height fluctuations are required to form contacts 
if the surface-surface separation is small. This suggests that the height of the contact cluster should 
have an impact on its fluctuations. Assuming a stable contact cluster exists, 
we calculate this height as

\begin{equation}
\label{eq:contact_height}
 H_c = \langle \textbf{h}\rangle_{\Omega_c} - \langle \textbf{h}\rangle_{\Omega_{\overline{c}}},
\end{equation}

\noindent
where the first average is over the domain which is in contact ($\Omega_c$),
and the second average is over the domain which is out of contact ($\Omega_{\overline{c}}$).

From the bottom row of Fig.~\ref{fig:fluctuations_rho} we observe 
only small changes in $H_c$ as the pressure is increased.
The reason for this is that even if the initial surface-surface separation is 
lower at higher pressures, once a stable contact is formed,
we are in practice locked into contact for the rest of the simulation.
From here on the system dissolves the part of the surface which is out of 
contact in order to grow the contact cluster to the optimal equilibrium height $H_c$ and size $\rhoWV$.
These two quantities are balanced since increasing $H_c$ increases the energy cost of growing new pillars to 
increase $\rhoWV$ (more particles are needed).

We also observe jumps in the contact fluctuations in Fig.~\ref{fig:fluctuations_rho}. 
However, looking at Fig.~\ref{fig:fluctuations_n},
where we plot the fluctuations in the number of gained and broken bonds separately, 
it is clear that these jumps are caused by the jumps in $\rhoWV$ due to transitions to a different contact favoring pressure level.

\subsection{Out-of-equilibrium systems}
\label{sec:out_of_equilibrium}

Here we will give a qualitative description of the typical behavior of the system out of equilibrium.

When the system has converged to equilibrium at some concentration $c_\mathrm{eq}$,
a supersaturation $\Omega\equiv c/c_\mathrm{eq} - 1$  can be introduced by 
setting the concentration to 

\begin{equation}
c_\Omega(t) = c_\mathrm{eq}(\Omega + 1),
\end{equation}

\noindent
and keeping it constant regardless of how many particles are dissolved or deposited.

\begin{figure*}[ht]
\begin{center}
 \includegraphics[width=1.0\textwidth]{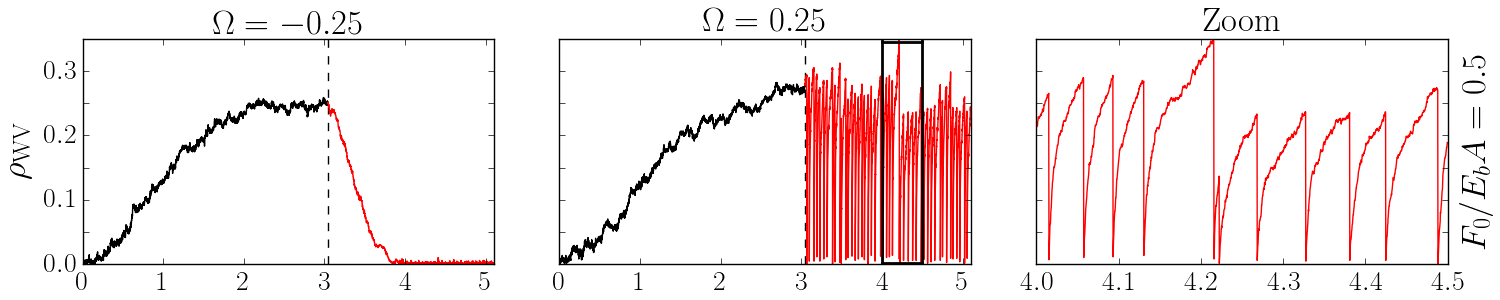} \\
 \includegraphics[width=1.0\textwidth]{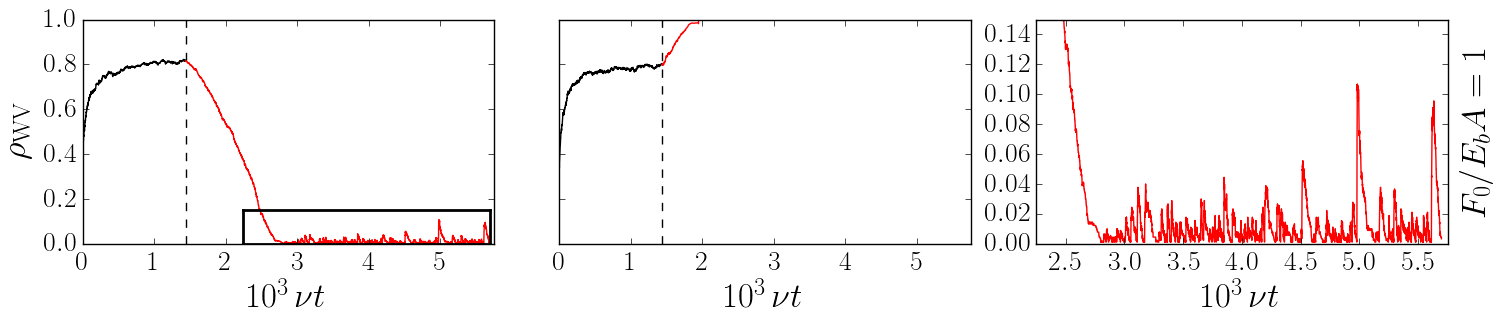} 
  \caption{Contact density $\rhoWV$ as a function of time for applied pressures $F_0/E_bA=0.5$ (top row)
  and $F_0/E_bA = 1$ (bottom row) using $\sigma_0 = 1$ and $E_b/kT = 1$. 
  At the point of the stippled vertical line, a constant saturation level $\Omega$ is applied. 
  Prior to this point the system has been equilibrated by fixing the effective number of particles. 
  The left column shows results for a dissolving system ($\Omega < 0$), 
  the middle column shown results for a growing system, and the right column shows
  zoom-ins of the black squares in the respective row. 
  Dissolution at low pressures is simply a decay of the initial contact cluster, and growth at high pressures simply results
  in the two surfaces merging completely ($\rhoWV=1$). 
  Growth at low pressures (top zoom) periodically causes the confining surface to separate from the crystal surface.
  For dissolution at high pressures (bottom zoom) we observe an initial stage where the contact cluster dissolves,
  after which we observe the contact density to spike every now and then.}
  \label{fig:noneq_stuff}
 \end{center}
\end{figure*}

Figure \ref{fig:noneq_stuff} shows the four typical behaviors the system possesses out of equilibrium.
For dissolving systems (left column where $\Omega < 0$) we see that if the pressure is low (top row), 
the equilibrium contacts simply dissolve and we end up with minor fluctuations and no stable contacts between the surfaces. 
Dissolution at high pressures (bottom row zoom-in) causes the initial large equilibrium contact to dissolve steadily as well, however, due to the high pressure,
the surfaces never decouple completely. 
This causes $\rhoWV$ to spike in value whenever the layer closest to the confining surface has been dissolved completely, 
since this enables the confining surface to snap down to the next layer.

For growth (middle column where $\Omega > 0$), we see that if the pressure is too high (bottom row), the surfaces simply grow into a perfect contact at $\rhoWV = 1$.
This occurs because the energy cost of breaking the contacts are very high due to the combined effect of a large number of contacts and a high pressure. 
This makes it so that being at rest on the crystal surface is the only mechanical equilibrium for the confining surface [condition 2(a) from Sec.~\ref{sec:mech_eq} always holds].

For growth at low pressures (center of the top row), 
we see that the that the surfaces repeatedly separate,i.e.~,$\rhoWV$ drops to 0,
after which it grows rapidly as particles are favored to stick in the newly formed cavity between the existing contact cluster
and the recently displaced confining surface. This newly formed contact has the same shape as the equilibrium contact. When the healing
of the broken contact has stagnated, the crystal surface starts to rise again, and the cycle repeats. 
This produces the jagged profile for growth at low pressures shown in the top zoom-in of Fig.~\ref{fig:noneq_stuff}.
Each repeating cycle lifts the confining surface one lattice unit.

This repeating separation occurs because
at low pressures the initial contact is small,
such that the system can build up enough repulsive energy to enable the mechanical equilibrium at $h_\lambda$ 
where the confining surface is not resting on the crystal surface.
When $h_\lambda$ is enabled, 
whether the surfaces separate is controlled by the bond breaking criteria 
from Sec.~\ref{sec:mech_eq}. 

The period of this cycle should therefore depend on two things: how quickly $h_\lambda$ can be enabled,
which is controlled by the supersaturation,
and the rate of breaking the bonds, which does not depend on the supersaturation, but 
on the attempt frequency (which we have set to every $A$ cycles).  
A larger pressure makes it harder to enable $h_\lambda$ and to break the bonds, thus it should increase the period.
If we increase the attempt frequency, the surfaces would separate sooner after $h_\lambda$ is enabled,
and if we decease it, the separation would happen later, until a point where the surfaces are able to merge completely before they are able to separate.

\section{Discussions and Conclusions}
\label{sec:Discussions}

The results suggest that the force of crystallization~\cite{Weyl1959}, which has been observed and studied in various early 
experiments~\cite{G.F.Becker1916, Taber1916} as well as new~\cite{Royne2012},
can cause a lifting of the confining surface even when there are stable contact clusters between the surfaces.
This means that we could have potentially large and irregular surface variations in such a system simply due to attractive interactions 
promoting contacts between the surfaces. This fact is in agreement with recent experimental observations~\cite{Royne2012}.
For dissolution at high pressures we found the surfaces to possess a continuous state of 
contact after the dominant equilibrium contact cluster dissolved.
Hence the results also suggest that pressure solution might occur with parts of the surfaces being in contact.
This is in agreement with experiments observations~\cite{Dysthe2002}.

This means that we by including attractive surface-surface interactions 
made the original model~\cite{Hogberget2016} go from producing solely flat interfaces, 
to producing structural roughnesses in the limits of low pressured growth and high pressured
dissolution. This suggests that our hypothesis stating that attractive surface-surface interactions
are essential mechanisms in confined crystallization is feasible.

Generally we observe a dominant equilibrium contact cluster forming between the surfaces from an initial state with no such contact.
The pressure and temperature dependency of the contact cluster size indicate that it is limited by thermal surface fluctuations (roughness).
If the fluctuations are insufficient then no stable contacts are formed.
During equilibration we recognize known concepts from kinetic limited growth theory such as 
primary and secondary nucleation stages, 
coalescence and Ostwald ripening. These effects are known to appear in island nucleation and step growth~\cite{Saito1996, pimpinelli},
and their presence suggests that the kinetics are properly treated in our model.

Since the initial surface-surface separation depends smoothly on the external pressure, 
and the thermal fluctuations depend smoothly on 
the temperature, the fact that fluctuations limit the contact size
tells us that the coexistence of the repulsive and attractive interactions is stable
not only for a selective few parameters, but for all.

The fact that certain pressure levels promote the formation of contacts due to the
initial surface-surface separation being such that formed contacts are especially hard to dissolve, 
is phenomenologically similar to oscillatory hydration forces~\cite{Israelachvili2011i}.
This shows that the model produces effects known to be associated with confined surfaces.

The size of the contact cluster relative to the system size, which we refer to as the contact density $\rhoWV$,
was found to be the key parameter deciding whether the equilibrium contact cluster was shaped as an island, a band or a pit,
as well as governing the stability of the contact in time (e.g.~the fluctuations and mobility).
The stability regions of these shapes were in excellent agreement with theoretical predictions.
This demonstrates that the model can be used to study details regarding both when and how contacts form between surfaces. 

Possible extensions of the model include introducing a Hamaker constant~\cite{Israelachvili2011i} in the attractive interaction to model 
a different material in the confining surface, and using the Eyring equation~\cite{Eyring1935} for the rates (i.e. calculate energy barriers),
such that a parameter for the bond breaking frequency would not be necessary. Moreover, a Lennard-Jones potential~\cite{Frenkel200223} could be 
used between the surfaces instead of the pure Van der Waals term we have used, which would remove the need to distinguish
between the surfaces resting on one another and being separated. Using discrete solute particles would enable the 
study of diffusion limited systems~\cite{paper2}. Elastic interaction could also be added to the surfaces~\cite{Lam2002, Lam2010, Russo2006, Schulze2011}.

Details aside, it is fascinating how much interesting physics came out of simply extending the previous model~\cite{Hogberget2016}
by counting the confining surface as a neighbor.
This leads us to believe that there is something simple yet fundamentally correct 
with our description of how attractive and repulsive forces work together in a confined system.

\begin{acknowledgments}
  This study was supported by the Research Council of Norway through the project ``Nanoconfined crystal growth and dissolution'' (No. 222386). 
  We acknowledge support from the Norwegian High Performance Computing (NOTUR) network through the grant of machine access.
\end{acknowledgments}

\appendix

\bibliography{paper3}
\bibliographystyle{apsrev4-1}

\end{document}